\documentclass[12pt]{article}
\usepackage{amsmath}
\usepackage{graphicx,psfrag,epsf}
\usepackage{enumerate}
\usepackage{multirow}
\usepackage[font=footnotesize]{caption}
\usepackage{subcaption}
\usepackage{rotating}
\usepackage{mwe}
\usepackage{mathtools}
\usepackage{amssymb}
\usepackage{diagbox}
\usepackage{url}
\usepackage{setspace}
\usepackage{natbib}
\setcitestyle{authoryear, aysep={,}, open={(},close={)}}

\newtheorem{theorem}{Theorem}
\newtheorem{lemma}{Lemma}

\newtheorem{proposition}{Proposition}

\newcommand{\blind}{0}
\DeclareMathOperator*{\argmin}{arg\,min}

\usepackage{algorithm}
\usepackage{algorithmic}

\begin{document}

\def\spacingset#1{\renewcommand{\baselinestretch}%
{#1}\small\normalsize} \spacingset{1}


\if0\blind
{
  \title{\bf  Functional Adaptive Double-Sparsity Estimator for Functional Linear Regression Model with Multiple Functional Covariates}
  \author{Cheng Cao \hspace{.2cm}\\
    School of Data Science, City University of Hong Kong \vspace{.2cm} \\
    Jiguo Cao \\
    Department of Statistics and Actuarial Science, Simon Fraser University \vspace{.2cm}  \\
    Hailiang Wang \\
    School of Design, The Hong Kong Polytechnic University \vspace{.2cm}  \\
    Kwok-Leung Tsui \\
    Grado Department of Industrial and Systems Engineering, \\ 
    Virginia Polytechnic Institute and State University \vspace{.2cm} \\
    Xinyue Li$^{*}$ \\
    School of Data Science, City University of Hong Kong}
  \maketitle
} \fi

\if1\blind
{
  \bigskip
  \bigskip
  \bigskip
  \begin{center}
    {\LARGE\bf Title}
\end{center}
  \medskip
} \fi

\let\thefootnote\relax\footnotetext{$^{*}$Corresponding author: Xinyue Li, Email: xinyueli@cityu.edu.hk}

\bigskip
\begin{abstract}
Sensor devices have been increasingly used in engineering and health studies recently, and the captured multi-dimensional activity and vital sign signals can be studied in association with health outcomes to inform public health. The common approach is the scalar-on-function regression model, in which health outcomes are the scalar responses while high-dimensional sensor signals are the functional covariates, but how to effectively interpret results becomes difficult. In this study, we propose a new Functional Adaptive Double-Sparsity (FadDoS) estimator based on functional regularization of sparse group lasso with multiple functional predictors, which can achieve global sparsity via functional variable selection and local sparsity via zero-subinterval identification within coefficient functions. We prove that the FadDoS estimator converges at a bounded rate and satisfies the oracle property under mild conditions. Extensive simulation studies confirm the theoretical properties and exhibit excellent performances compared to existing approaches. Application to a Kinect sensor study that utilized an advanced motion sensing device tracking human multiple joint movements and conducted among community-dwelling elderly demonstrates how the FadDoS estimator can effectively characterize the detailed association between joint movements and physical health assessments. The proposed method is not only effective in Kinect sensor analysis but also applicable to broader fields, where multi-dimensional sensor signals are collected simultaneously, to expand the use of sensor devices in health studies and facilitate sensor data analysis.

\end{abstract}

\noindent%

\spacingset{1.45}

\section{Introduction}
\label{sec:intro}

Sensor devices have been increasingly used in engineering and health studies recently, and the captured multi-dimensional activity and vital sign signals can be studied in association with health outcomes to inform public health. The common approach is the scalar-on-function regression model, in which health outcomes are the scalar responses while high-dimensional sensor signals are the functional covariates, but how to effectively interpret results becomes difficult. In this study, we propose a new Functional Adaptive Double-Sparsity (FadDoS) estimator based on functional regularization of sparse group lasso with multiple functional predictors, which can achieve global sparsity via functional variable selection and local sparsity via zero-subinterval identification within coefficient functions. We prove that the FadDoS estimator converges at a bounded rate and satisfies the oracle property under mild conditions. Extensive simulation studies confirm the theoretical properties and exhibit excellent performances compared to existing approaches. Application to a Kinect sensor study that utilized an advanced motion sensing device tracking human multiple joint movements and conducted among community-dwelling elderly demonstrates how the FadDoS estimator can effectively characterize the detailed association between joint movements and physical health assessments. The proposed method is not only effective in Kinect sensor analysis but also applicable to broader fields, where multi-dimensional sensor signals are collected simultaneously, to expand the use of sensor devices in health studies and facilitate sensor data analysis.

\begin{figure}[!b]
\centering
\begin{subfigure}{.3\textwidth}
  \centering
  \includegraphics[width=1.2\textwidth, height=1\textwidth]{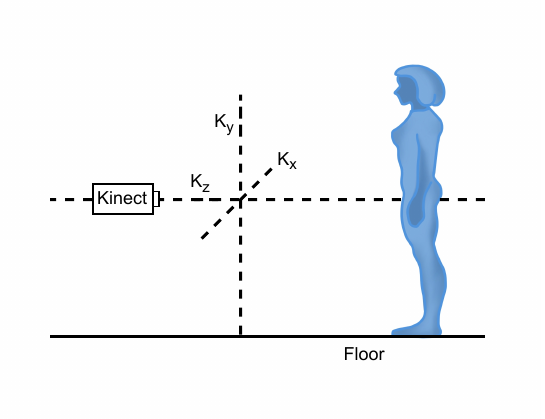}
  \caption{}
\end{subfigure}
\begin{subfigure}{.3\textwidth}
  \centering
  \includegraphics[width=1\textwidth, height=1\textwidth]{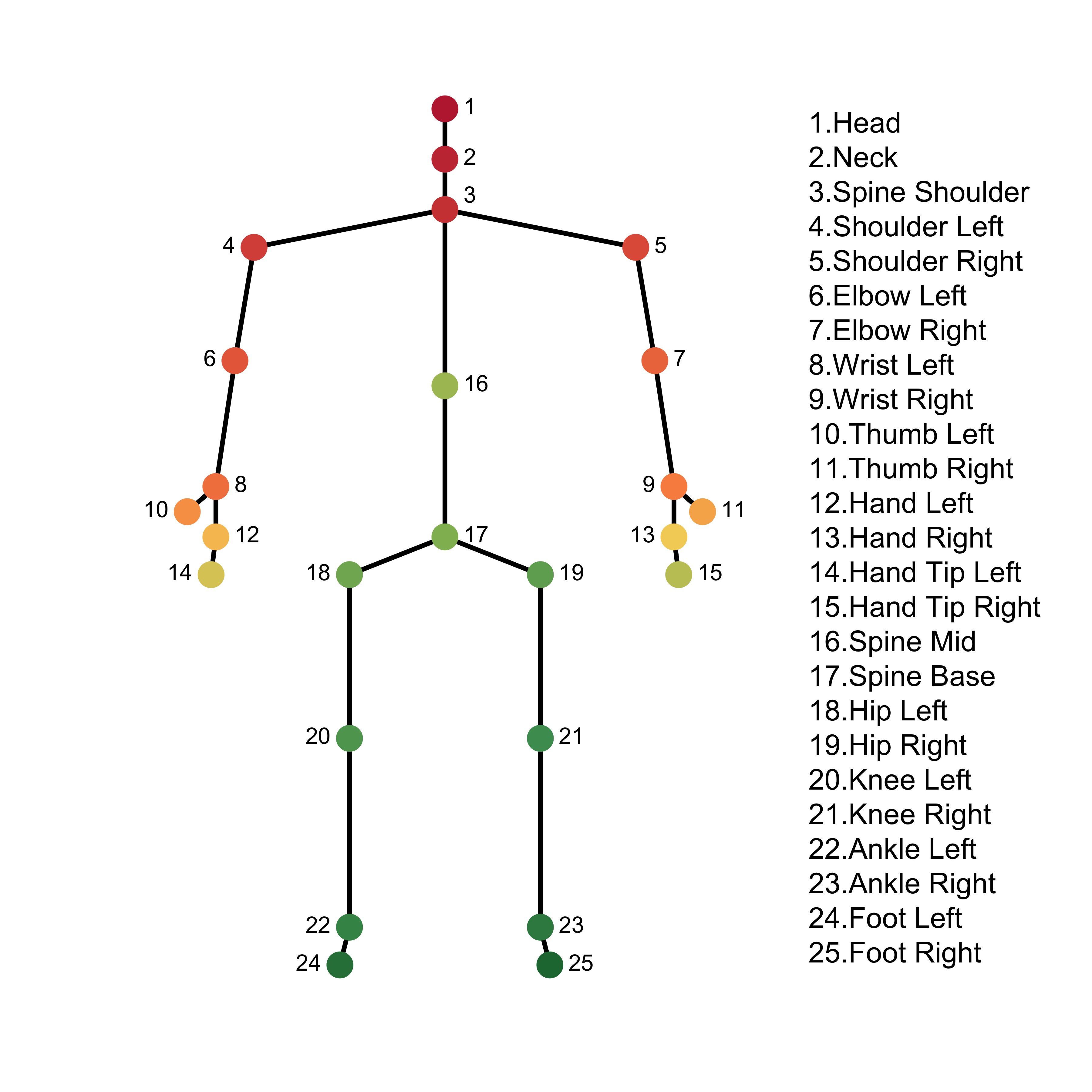}
  \caption{}
\end{subfigure}
\begin{subfigure}{.3\textwidth}
  \centering
  \includegraphics[width=1\textwidth, height=1\textwidth]{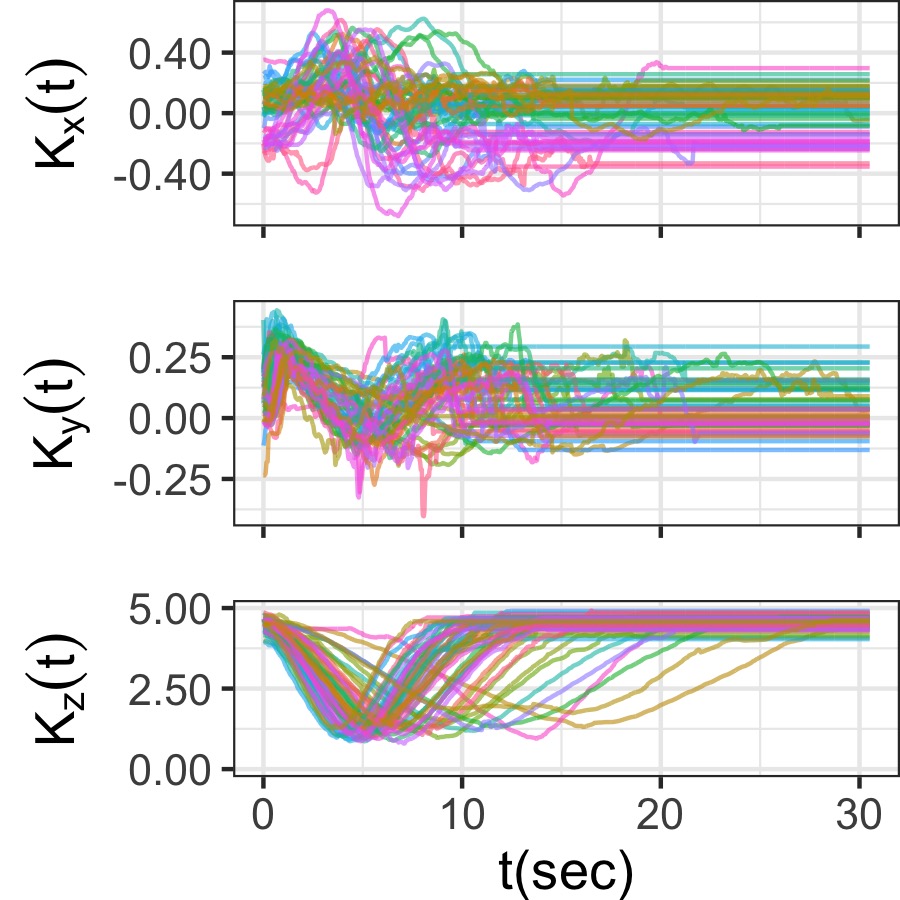}  
  \caption{}
\end{subfigure}
\caption{(a) The setup of Kinect sensor. (b) The 25 joints of human skeleton captured by Kinect sensor. (c) The three-dimensional displacement of the sacrum (joint 17: Spine Base) for all participants in Timed Up and Go (TUG) Test. Each curve represents time-dependent sacrum position of a participant during the entire test. Due to the different duration of finishing, we use the last observation to pad the shorter signals till all signals have equal time domain. The participant ID is color coded. Top panel: horizontal direction. Middle panel: vertical direction. Bottom panel: walking direction.}
\label{fig:setup}
\end{figure}

However, the analysis of the complex high-dimensional Kinect data, same as other types of multi-dimensional sensor device data, remains challenging. Given a large number of functional covariates and measured scalar responses, we consider scalar-on-function regression model with multiple functional covariates (\citet{ramsay}). The estimates of the coefficient functions can yield insights into the detailed associations between functional covariates and the outcome over the time domain. To enhance the interpretability of the complex model, it is crucial to conduct functional variable selection and further obtain sparse coefficient estimates, which we refer to as global sparsity and local sparsity respectively, to obtain meaningful results and relevant interpretation of the associations.

Several estimators targeting scalar-on-function regression model with multiple functional covariates address global sparsity to achieve variable selection (\citet{pannu1, collazos, cheng}). \cite{gertheiss} proposed a sparsity-smoothness estimator to obtain a parsimonious model based on functional generalization of group lasso penalty (\citet{meier}), while \cite{matsui} and \cite{mingotti} suggested alternative functional regularizations. However, these estimators without concerning local sparsity can only facilitate the interpretability with respect to variable selection. With increasing attention to coefficient interpretability, more methods aim to preserve local sparsity of functional coefficient (\citet{james, zhou}). The sparse estimation of functional coefficient is reformulated to parametric variable selection in basis function representation of linear regression with various regularizations. \cite{wangh} and \cite{lin} proposed an estimator based on the group bridge and functional generalization of SCAD penalty, respectively. These approaches can identify zero subregions with nice asymptotic properties, but both primarily target a single functional covariate. \cite{tu} and \cite{tu1} extended the group bridge penalty to multiple functional covariates with binary and functional responses rather than scalar responses, while the suboptimal local solution due to the non-convex penalty could be a concern.

In this paper, we propose a novel functional adaptive double-sparsity (FadDoS) estimator, which can achieve global sparsity via functional variable selection and local sparsity via sparse coefficient estimation simultaneously in scalar-on-function regression models with multiple functional covariates. The combination of global and local sparsity is termed double-sparsity. The proposed estimator achieves double-sparsity through functional generalization of sparse group lasso and adaptive penalization. By solving an optimization problem that incorporates regularization and smoothing splines in a single convex objective function, we can provide the estimates of coefficient functions with nice double-sparsity and smoothness control. The benefit of convexity is the guarantee of converging to a local minimum during model fitting. 

The major contribution of our work is three-fold. First, we defined an important property, double-sparsity, of functional linear regression with multiple functional covariates, which leads to sparse and consistent functional estimates when not all functional covariates over the entire time domain are associated with the scalar response. Second, we provided a one-stage estimation procedure for the FadDoS estimator to achieve double-sparsity, which is computationally efficient and easy to implement. Third, the accuracy of the estimation is theoretically stable and solid as we prove that the FadDoS estimator converges at an optimal rate and enjoys oracle properties. Our method can remarkably enhance model parsimony and interpretability for high-dimensional sensor device data that existing methods in the literature lack. As an example, the FadDoS estimator can be applied to complex Kinect sensor data to effectively examine the association between joint movements and mobility, based on which automated assessment tools can be further developed for health monitoring and early intervention.

The rest of this paper is organized as follows. In Section 2, we introduce the double-sparsity structure of the scalar-on-function regression model with multiple functional covariates and computation of the FadDoS estimator. Section \ref{sec:theo} presents asymptotic properties of the proposed estimator. Extensive simulation studies are conducted in Section \ref{sec:simu} to evaluate the performance of our estimator and compare the results with existing approaches. Section \ref{sec:app} describes the application for the Kinect elderly study. Conclusion and discussion are in Section \ref{sec:conc}. All code for model implementation and simulation is available in the supplementary material and at \url{https://github.com/Cheng-0621/FadDoS}.

\section{Methodology}
\label{sec:meth}

\subsection{Global and Local Sparsity in Scalar-on-function Regression Model with Multiple Functional Covariates}

Suppose that $Y_{i}$ be the scalar response and $X_{ij}(t) \in L^{2}(\mathcal{T})$ be the $j$th functional covariate for subject $i$ observed at time $t$ in domain $\mathcal{T}$. The functional linear model with multiple functional covariates is thus
\begin{equation}
Y_{i}  =\mu + \sum_{j=1}^{J}\int_{\mathcal{T}}X_{ij}(t)\beta_{j}(t)dt + \epsilon_{i}, \ \ i=1,\dots,n,
\label{flm}
\end{equation}
where $\mu$ is the intercept term and $ \epsilon_{i} \sim N(0, \sigma^{2})$. The smooth coefficient function $\beta_{j}(t)$ provides a time-dependent contribution of $X_{ij}(t)$ to the corresponding response $Y_{i}$. 

We use the following notation: for all measurable functions $f$ whose absolute value raised to the $q$-th power has a finite integral, which is $||f||_{q} = (\int_{\mathcal{T}} |f(t)|^{q}dt)^{1/q} < \infty$, and the supremum norm of a function $f$ is defined as $||f||_{\infty} = \sup_{t \in \mathcal{T}}|f(t)|$. The double-sparsity structure of the functional linear model \eqref{flm} contains both global sparsity and local sparsity. Mathematically, the global sparsity is characterized by the set of indices for which functional covariates have no contribution to the response over the entire domain, i.e. $\{j \in \{1,\dots, J\}: \beta_{j}(t) = 0, \text{for all} \ t \in \mathcal{T}\}$, while for those functional covariates not belonging to this set, the local sparsity is depicted by the partition of zero and nonzero subregion of the corresponding coefficient functions. Suppose a subregion $I \subset \mathcal{T}$, the zero subregion is defined $N_{0}(\beta_{j})=\{I \subset \mathcal{T}:  \beta_{j}(t)=0,  \text{for all} \ t \in I\}$ and thus the nonzero subregion $N_{1}(\beta_{j})$ is the complement of $N_{0}(\beta_{j})$. It implies that within the subinterval $N_{0}(\beta_{j})$, the functional covariate has no effect on the response. With the double-sparsity structure, important functional predictors can be selected and nonzero subintervals can be further identified.  

\subsection{Functional Sparse Group Lasso Penalization}

To achieve both global and local sparsity, we propose a novel regularization method based on functional generalization of sparse group lasso (fSGL), where SGL is initially proposed by \cite{friedman} and \cite{simon}. The penalty function is
\begin{equation}
  P_{\lambda_{1}, \lambda_{2}, \varphi}(\boldsymbol{\beta}) =  \lambda_{1}||\boldsymbol{\beta}||_{1}  + \lambda_{2}\sum_{j=1}^{J}(||\beta_{j}||^{2}_{2} + \varphi||\mathcal{D}^{m}\beta_{j}||^{2}_{2})^{1/2},
  \label{eq:panalizatio:fsgl}
\end{equation}
where $\boldsymbol{\beta}=(\beta_{1}, \dots, \beta_{J})^{T}$ is the vector of coefficient functions, $||\boldsymbol{\beta}||_{1} = \sum_{j=1}^{J}||\beta_{j}||_{1}$ with $||\beta_{j}||_{1} = \int_{\mathcal{T}} |\beta_{j}(t)|dt$, $||\beta_{j}||^{2}_{2} = \int_{\mathcal{T}}|\beta_{j}(t)|^2dt$ and $||\mathcal{D}^{m}\beta_{j}||^{2}_{2} = \int_{\mathcal{T}} |\partial^{m}\beta_{j}(t)/\partial t^{m}|^2dt$ such that $m \leq d$, where $d$ is the degree of polynomial functions, $\lambda_{1}, \lambda_{2}, \varphi \geq 0$. The additive penalties contain two convex regularizations. The functional generalization of $\ell_{1}$ (functional $\ell_{1}$) penalty regularizes the magnitude of local sparseness, while functional generalization of $\ell_{1,2}$ (functional $\ell_{1,2}$) penalty manages global sparseness. Within functional $\ell_{1,2}$ penalty, the roughness penalty $\varphi||\mathcal{D}^{m}\beta_{j}||_{2}$ is employed to apply penalization to the $m$-th order derivative of the $j$th coefficient function to control the smoothness. It is worth noting that achieving this goal is also possible by placing $\varphi||\mathcal{D}^{m}\beta_{j}||_{2}$ outside the square root, whereas our functional $\ell{1,2}$ penalty offers an advantage of computational efficiency for solving a simpler optimization problem, same as discussed in \citet{meier}. Thus, the minimizer of mean squared loss function regularized by fSGL in Equation \eqref{eq:panalizatio:fsgl} is called functional double-sparsity (FDoS) estimator.

However, one drawback of this regularization method is that the magnitude of penalization is the same for all functional covariates in the model, which may result in either insufficient suppression of zero functional estimates or underestimation of nonzero functional estimates. Theoretically, the equal degrees of penalization fail to satisfy the oracle property suggested in \cite{fan}, unless a proper adaptive method is applied to enable importance re-weighting of different coefficient functions (\citet{zou}).

\subsection{Functional Adaptive Double-Sparsity Estimator}

We further enhance the fSGL penalization with adaptive weights to allow for flexible regularization to different coefficient functions. The adaptive penalty function is
\begin{equation}
  P_{\lambda_{1}, \lambda_{2}, \varphi}(\boldsymbol{\beta}) =  \lambda_{1}\sum_{j=1}^{J}w^{(1)}_{j}||\beta_{j}||_{1}  + \lambda_{2}\sum_{j=1}^{J}w^{(2)}_{j}(||\beta_{j}||^{2}_{2} + \varphi||\mathcal{D}^{m}\beta_{j}||^{2}_{2})^{1/2}, 
  \label{eq:panalizatio:fadsgl}
\end{equation}
where $w^{(1)}_{j} = ||\check{\beta}_{j}||_{1}^{-a}$ and $w^{(2)}_{j}= ||\check{\beta}_{j}||_{2}^{-a}$ are known nonnegative weights and $\check{\beta}_{j}$ is the initial estimator, $a> 0$ is an adjustment of the adaptive weights and usually $a = 1$.

Intuitively, the adaptive weights can incorporate prior insights into functional covariates: a larger value implies less importance in the model. The adaptive approach can effectively penalize zero functions to reduce false negatives without over-regularizing nonzero functions. Additionally, the adaptive weights in Equation \eqref{eq:panalizatio:fadsgl} are predictor-wise rather than subregion-wise, i.e., $||w^{(1)}_{j}\beta_{j}||_{1}$ with known nonnegative weight function $w^{(1)}_{j}(t) = |\check{\beta}_{j}(t)|^{-a}$. We do not choose subregion-wise weight because it is computationally intensive and more importantly, prone to errors, as the coefficient estimation is largely affected by choices of initial estimators. While the adaptive estimation allows for different initial estimators, we choose the smoothing spline estimator as in \cite{cardot} with the second-order differential operator in the penalty derived from functional generalization of ridge penalization with generally no sparse solution, to avoid explosion to infinity when calculating the reciprocal of the norm even for those less significant functional covariates. 

Given the adaptive fSGL penalization in Equation \eqref{eq:panalizatio:fadsgl}, we propose the FadDoS estimator, which minimizes the penalized mean squared loss defined by
\begin{equation}
\begin{aligned}
 L_{n}(\boldsymbol{\beta}, \mu) = & \frac{1}{2}\sum_{i=1}^{n} \Big[ Y_{i} - \mu - \sum_{j=1}^{J}\int_{\mathcal{T}}X_{ij}(t)\beta_{j}(t)dt \Big]^{2} + \\ 
 & \lambda_{1}\sum_{j=1}^{J}w^{(1)}_{j}||\beta_{j}||_{1}  + \lambda_{2}\sum_{j=1}^{J}w^{(2)}_{j}(||\beta_{j}||^{2}_{2} + \varphi||\mathcal{D}^{m}\beta_{j}||^{2}_{2})^{1/2}.
\end{aligned}
  \label{eq:obj-fadsgl-genlasso1}
\end{equation}
It is obvious that FDoS is a special case of FadDoS if fixing $w^{(1)}_{j} = w^{(2)}_{j} = 1$ for all $j=1,\dots, J$. Moreover, $m=2$ is commonly chosen and we adopt it for the rest of the paper.

The computational procedure of the FadDoS estimator is shown below. We first express the coefficient functions $\beta_{j}(t)$ with B-splines. Suppose that there are $M_{n}+1$ equally spaced knots $0=t_{j0}<t_{j1}<\dots<t_{jM_{n}}=T$ in the domain $\mathcal{T}$ and $d$ degree of polynomial functions, and let $\mathcal{S}_{j}$ be the linear space spanned by basis functions $\{B_{jk}(t): k=1,\dots,M_{n}+d\}$ in $\mathcal{T}$, we have $\beta_{j}(t) = \sum_{k=1}^{M_{n}+d} B_{jk}(t)b_{jk} = \boldsymbol{B}_{j}^{T}(t)\boldsymbol{b}_{j}$, where $\boldsymbol{B}_{j}(t)=(B_{j1}(t), \dots, B_{j(M_{n}+d)}(t))^{T}$, $\boldsymbol{b}_{j}=(b_{j1}, \dots , b_{j(M_{n}+d)})^{T}$. With the compact support property of B-splines, the basis function can characterize the sparseness of coefficient functions as it is a nonzero polynomial over no more than $d+1$ adjacent subintervals $[t_{j(r-d-1)}, t_{jr})$, $r \in \{1,\dots, M_{n}\}$, implying only $d+1$ basis functions have supports within a subinterval $[t_{j(r-1)}, t_{jr})$. 

Using B-spline basis functions, we reparameterize Equation \eqref{eq:obj-fadsgl-genlasso1} to facilitate computation. First, $||\beta_{j}||_{1}$ can be represented by basis coefficients given by property illustrated in \cite{deboor, mingotti}. Mathematically, let $\Delta_{n}=t_{jr}-t_{j(r-1)}$, by Riemann Integral and the B-spline property, we have 
\begin{equation*}
    ||\beta_{j}||_{1} = \int_{\mathcal{T}}|\beta_{j}(t)|dt \approx \Delta_{n}\sum_{r=1}^{M_{n}}|\beta_{j}(t_{jr})| \approx \Delta_{n}\sum_{k=1}^{M_{n}+d}|b_{jk}| = \Delta_{n}||\boldsymbol{b}_{j}||_{1}.
\end{equation*}
Hence, functional $\ell_{1}$ penalty term can be replaced by $\Delta_{n}\sum_{j=1}^{J}||\boldsymbol{b}_{j}||_{1}$. Additionally, the functional $\ell_{1,2}$ penalty term can be rewritten as $\sum_{j=1}^{J}(||\beta_{j}||^{2}_{2} + \varphi||\mathcal{D}^{2}\beta_{j}||^{2}_{2})^{1/2}=\sum_{j=1}^{J}(\boldsymbol{b}_{j}^{T}(\boldsymbol{\Phi}_{j}+\varphi\boldsymbol{\Omega}_{j})\boldsymbol{b}_{j})^{1/2}$, where $\boldsymbol{\Phi}_{j}$ is a $(M_{n}+d)\times(M_{n}+d)$ matrix with the $(p,q)$ element equal to $(\boldsymbol{\Phi}_{j})_{pq} = \int B_{jp}(t)B_{jq}(t)dt$, and $\boldsymbol{\Omega}_{j}$ is also a $(M_{n}+d)\times(M_{n}+d)$ matrix with the $(p,q)$ element equal to $(\boldsymbol{\Omega}_{j})_{pq} = \int B^{(2)}_{jp}(t)B^{(2)}_{jq}(t)dt$. In practice, with dense grids observed over the domain $\mathcal{T}$, we compute $(\boldsymbol{\Phi}_{j})_{pq} \approx \Delta_{n}\sum_{r} B_{jp}(t_{jr})B_{jq}(t_{jr})$ and $(\boldsymbol{\Omega}_{j})_{pq} \approx \Delta_{n}\sum_{r} B^{(2)}_{jp}(t_{jr})B^{(2)}_{jq}(t_{jr})$. Let $\boldsymbol{K}_{\varphi,j,\Delta_{n}}=(\boldsymbol{\Phi}_{j}+\varphi\boldsymbol{\Omega}_{j})/\Delta_{n}^{2}$, and we can use Cholesky decomposition to have $\boldsymbol{K}_{\varphi,j,\Delta_{n}} = \boldsymbol{L}_{j}\boldsymbol{L}_{j}^{T}$, where $\boldsymbol{L}_{j}$ is a $(M_{n}+d)\times(M_{n}+d)$ non-singular lower triangular matrix. By defining a new vector of coefficients $\tilde{\boldsymbol{b}}_{j} = \Delta_{n} \boldsymbol{L}_{j}^{T}\boldsymbol{b}_{j}$ and $\tilde{\boldsymbol{b}} = (\tilde{\boldsymbol{b}}_{1}^{T}, \dots, \tilde{\boldsymbol{b}}_{J}^{T})^{T}$, we obtain the equivalent matrix form of the objective function in Equation \eqref{eq:obj-fadsgl-genlasso1},
\begin{equation}
     L_{n}(\tilde{\boldsymbol{b}}, \boldsymbol{\mu})= \frac{1}{2}|| \boldsymbol{Y} - \boldsymbol{\mu} - \sum_{j=1}^{J}\widetilde{\boldsymbol{U}}_{j}\tilde{\boldsymbol{b}}_{j}||_{2}^{2}+ \lambda_{1}\sum_{j=1}^{J}w^{(1)}_{j}||(\boldsymbol{L}_{j}^{T})^{-1} \tilde{\boldsymbol{b}}_{j}||_{1} + \lambda_{2} \sum_{j=1}^{J}w^{(2)}_{j}||\tilde{\boldsymbol{b}}_{j}||_{2},   
\label{eq:obj-fsgl-genlasso1}     
\end{equation}
where $\boldsymbol{Y}=(Y_{1}, \dots, Y_{n})^{T}$, $\boldsymbol{\mu} = \mu\mathbf{1}_{n}$, $\mathbf{1}_{n} = (1,\dots,1)^{T}$ denotes a $n$-vector of ones. $\widetilde{\boldsymbol{U}}_{j} = \boldsymbol{U}_{j}(\boldsymbol{L}_{j}^{T})^{-1}/\Delta_{n}$ where $\boldsymbol{U}_{j}$ is an $n \times (M_{n}+d)$ matrix with entries $(\boldsymbol{U}_{j})_{ip} = \int_{\mathcal{T}} X_{ij}(t)B_{jp}(t)dt$ that can be approximated by $\Delta_{n}\sum_{r} X_{ij}(t_{jr})B_{jp}(t_{jr})$. The solution to the FadDoS estimator of $\hat{\beta}_{j}(t)$ is thus $\hat{\beta}_{j}(t) = \Delta_{n}^{-1}\boldsymbol{B}_{j}^{T}(\boldsymbol{L}_{j}^{T})^{-1}\tilde{\boldsymbol{b}}^{*}_{j}$ and the objective turns to estimate $\tilde{\boldsymbol{b}}^{*}_{j}$.

\subsection{Algorithm}

The algorithm for solving Equation \eqref{eq:obj-fsgl-genlasso1} was developed based on Alternating Direction Method of Multipliers (ADMM). ADMM is effective in lasso and group lasso, and has been recently extended to non-separable and non-smooth convex problems such as fused lasso, as it is powerful in leveraging the complicated structure of $\ell_{1}$ norm (\citet{boyd, lix, beer}). Here we choose ADMM because it does not require a differentiable objective function and at the same time supports decomposition to split the objective into solvable subproblems.

We first define two common operators: for $\lambda, \kappa >0$ and $y \in \mathbb{R}^{p}$, the soft thresholding operator $S_{1, \lambda/\kappa}(y) = \text{sgn}(y)(|y|-\lambda/\kappa)_{+}$ and $ S_{2, \lambda/\kappa}(y) = y/||y||_{2} (||y||_{2}-\lambda/\kappa)_{+}$. For ease of notation, we assume no intercept, i.e., $\mu = 0$ and consider $w^{(1)}_{j}=w^{(2)}_{j}=1$ for all $j$ to illustrate the main idea, which is equivalent to the FDoS estimator. The estimate of $\mu$ for the case $\mu \neq 0$ is shown at the end of the section. Let $\boldsymbol{D}$ be a block diagonal matrix, $\boldsymbol{D} = \text{diag}((\boldsymbol{L}_{1}^{T})^{-1}, (\boldsymbol{L}_{2}^{T})^{-1}, \dots, (\boldsymbol{L}_{J}^{T})^{-1})$, the optimization problem in Equation \eqref{eq:obj-fsgl-genlasso1} can be rewritten as
\begin{equation}
\begin{aligned}
 \min_{\tilde{\boldsymbol{b}}_{j} \in \mathbb{R}^{M_{n}+d}, j=1,\dots,J} \quad
   & \frac{1}{2}||\boldsymbol{Y}-\sum_{j=1}^{J}\widetilde{\boldsymbol{U}}_{j}\tilde{\boldsymbol{b}}_{j}||_{2}^{2} + \lambda_{1}||\boldsymbol{z}||_{1}  + \lambda_{2} \sum_{j=1}^{J}||\tilde{\boldsymbol{b}}_{j}||_{2}  \\
  \textrm{s.t.} \quad & \boldsymbol{z}= \boldsymbol{D}\tilde{\boldsymbol{b}}
\end{aligned}   
\end{equation}
and the corresponding augmented Lagrangian function $\mathcal{L}(\tilde{\boldsymbol{b}}, \boldsymbol{z}, \boldsymbol{u}) = \frac{1}{2}||\boldsymbol{Y}-\sum_{j=1}^{J}\widetilde{\boldsymbol{U}}_{j}\tilde{\boldsymbol{b}}_{j}||_{2}^{2} + \lambda_{1}||\boldsymbol{z}||_{1} + \lambda_{2} \sum_{j=1}^{J}||\tilde{\boldsymbol{b}}_{j}||_{2}  + \boldsymbol{u}^{T}(\boldsymbol{D}\tilde{\boldsymbol{b}} - \boldsymbol{z}) + \rho/2||\boldsymbol{D}\tilde{\boldsymbol{b}} - \boldsymbol{z}||_{2}^{2}$, where $\rho > 0$ is the prespecified augmented Lagrangian parameter, namely the step size to update the dual variable. The iteration scheme of ADMM consists of three sub-problems shown below. 

First, we deal with the $\tilde{\boldsymbol{b}}$-update at the $k$th iteration. Since the penalty terms are separable, we can update $\tilde{\boldsymbol{b}}^{k+1}_{j}$ parallelly at the $k$th iteration. Assuming other coefficients fixed, we define $\boldsymbol{r}_{(-l)} = \boldsymbol{Y} - \sum_{j \neq l} \widetilde{\boldsymbol{U}}_{j}\tilde{\boldsymbol{b}}^{k}_{j}$, $\boldsymbol{z}^{k}=\big((\boldsymbol{z}^{k}_{1})^{T}, \dots, (\boldsymbol{z}^{k}_{J})^{T}\big)^{T}$ and $\boldsymbol{u}^{k}=\big((\boldsymbol{u}^{k}_{1})^{T}, \dots, (\boldsymbol{u}^{k}_{J})^{T}\big)^{T}$, then
\begin{equation}
\begin{aligned}
    \tilde{\boldsymbol{b}}_{l}^{k+1}  = & \argmin_{\tilde{\boldsymbol{b}}_{l} \in \mathbb{R}^{M_{n}+d}} \frac{1}{2}||\boldsymbol{r}_{(-l)}-\widetilde{\boldsymbol{U}}_{l}\tilde{\boldsymbol{b}}_{l}||_{2}^{2} + \lambda_{2} ||\tilde{\boldsymbol{b}}_{l}||_{2}  + (\boldsymbol{u}^{k}_{l})^{T}((\boldsymbol{L}_{l}^{T})^{-1} \tilde{\boldsymbol{b}}_{l}-\boldsymbol{z}^{k}_{l}) + \\ & \rho/2||(\boldsymbol{L}_{l}^{T})^{-1} \tilde{\boldsymbol{b}}_{l} - \boldsymbol{z}^{k}_{l}||_{2}^{2}.
\end{aligned}
\end{equation}
To further solve the optimization problem, we employ a linearization technique to approximate the quadratic term efficiently (\citet{wangx}), and details of derivation are provided in the supplementary file. We let $\widehat{\boldsymbol{U}}_{l} = \big(\widetilde{\boldsymbol{U}}_{l}^{T}, \rho^{1/2}\boldsymbol{L}_{l}^{-1}\big)^{T}$ and $\hat{\boldsymbol{r}}_{(-l)}^{k} = \big(\boldsymbol{r}_{(-l)}^{T}, \rho^{1/2}(\boldsymbol{z}^{k}_{l} + \boldsymbol{u}^{k}_{l}/\rho\big)^{T})^{T}$, and thus the $\tilde{\boldsymbol{b}}$-update solution is computed by the second operator such that $\tilde{\boldsymbol{b}}_{l}^{k+1} = S_{2, \lambda_{2}/\nu_{l}}(\tilde{\boldsymbol{b}}_{l}^{k} - \widehat{\boldsymbol{U}}_{l}^{T}\big(\widehat{\boldsymbol{U}}_{l}\tilde{\boldsymbol{b}}_{l}^{k}-\hat{\boldsymbol{r}}_{(-l)}^{k} \big)/\nu_{l})$,where $\nu_{l} > 0$ is the proximal parameter that controls the proximity to $\tilde{\boldsymbol{b}}_{l}^{k}$. Note that $\nu_{l}$ is required to be larger than the spectral radius of $\widetilde{\boldsymbol{U}}_{l}^{T}\widetilde{\boldsymbol{U}}_{l} + \rho\mathbf{I}$ for convergence (\citet{wangx}). 

Second, the $\boldsymbol{z}$-update at the $k$th iterations is
\begin{equation}
        \boldsymbol{z}^{k+1} = \argmin_{\boldsymbol{z} \in \mathbb{R}^{J\times(M_{n}+d)}} \lambda_{1}||\boldsymbol{z}||_{1} + (\boldsymbol{u}^{k})^{T}(\boldsymbol{D}\tilde{\boldsymbol{b}}^{k+1} - \boldsymbol{z}) + \rho/2||\boldsymbol{D}\tilde{\boldsymbol{b}}^{k+1} - \boldsymbol{z}||_{2}^{2}.
\end{equation}
The $\boldsymbol{z}$-update solution is computed by the first operator such that $\boldsymbol{z}^{k+1} = S_{1, \lambda_{1}/\rho}(\boldsymbol{D}\tilde{\boldsymbol{b}}^{k+1} + \boldsymbol{u}^{k}/\rho)$. Details of derivation are given in the supplementary file.

Third, the dual variable update at the $k$th iteration is $\boldsymbol{u}^{k+1} = \boldsymbol{u}^{k} + \rho( \boldsymbol{D}\tilde{\boldsymbol{b}}^{k+1} - \boldsymbol{z}^{k+1})$.

\begin{spacing}{1} 
\begin{algorithm}[!t]
\caption{ADMM FDoS}
\begin{algorithmic}

\REQUIRE Choose tuning parameters $\lambda_{1}$, $\lambda_{2}$, $\varphi > 0$; step size $\rho > 0$; B-splines parameters $M_{n}$ and $d$; 
\ENSURE $\tilde{\boldsymbol{b}}^{0}$, $\boldsymbol{z}^{0}$, $\boldsymbol{u}^{0}$

\FOR{$k=1,2,\dots,$} 

\FOR{$l=1,2,\dots,J$}

\STATE Let $\nu_{l}$ be the spectral radius of  $\widetilde{\boldsymbol{U}}_{l}^{T}\widetilde{\boldsymbol{U}}_{l} + \rho\mathbf{I}$ multiplied by 5
\STATE Compute $\tilde{\boldsymbol{b}}^{k+1}_{l} = S_{2, \lambda_{2}/\nu_{l}}(\tilde{\boldsymbol{b}}_{l}^{k} - \widehat{\boldsymbol{U}}_{l}^{T}\big(\widehat{\boldsymbol{U}}_{l}\tilde{\boldsymbol{b}}_{l}^{k}-\hat{\boldsymbol{r}}_{(-l)}^{k} \big)/\nu_{l})$

\ENDFOR
\STATE Compute $\boldsymbol{z}^{k+1} = S_{1, \lambda_{1}/\rho}(\boldsymbol{D}\tilde{\boldsymbol{b}}^{k+1} + \boldsymbol{u}^{k}/\rho)$
\STATE Compute $\boldsymbol{u}^{k+1} = \boldsymbol{u}^{k} + \rho( \boldsymbol{D}\tilde{\boldsymbol{b}}^{k+1} - \boldsymbol{z}^{k+1})$

\ENDFOR \ \ until stopping criteria with a pre-specified tolerance $\epsilon_{\text{tol}}$ such that  $$||\tilde{\boldsymbol{b}}^{k}- \tilde{\boldsymbol{b}}^{k-1}||_{2}/\max\{||\tilde{\boldsymbol{b}}^{k}||_{2},1\} \leq \epsilon_{\text{tol}}$$ is reached

\COMMENT \ \ $\boldsymbol{z}^{*}$ and $\tilde{\boldsymbol{b}}^{*} = \boldsymbol{D}^{-1}\boldsymbol{z}^{*}$
\end{algorithmic}
\end{algorithm}
\end{spacing} 

The full algorithm is described in Algorithm 1. When $\mu \neq 0$, we can estimate $\hat{\mu}^{k+1}$ after $\boldsymbol{z}$-update at the $k$th iteration by computing $\hat{\mu}^{k+1} = n^{-1}\mathbf{1}_{n}^{T}(\boldsymbol{Y}-\widetilde{\boldsymbol{U}}\boldsymbol{D}^{-1}\boldsymbol{z}^{k+1})$ until the stopping criteria is reached. In addition, the set of three tuning parameters is determined by cross-validation, assuming $\rho$ prespecified,  $M_{n}$ and $d$ fixed.

\section{Asymptotic Properties}
\label{sec:theo}

We study the asymptotic properties of our proposed estimators in the fixed dimensional models, where the number of functional covariates does not depend on the sample size. Under some regularity conditions, we show that the proposed estimator converges to the true coefficient function in a bounded rate. As established in prior work (\citet{zou, nardi, poignard}), lasso-type estimators including group lasso and sparse group lasso cannot achieve the oracle property without appropriate adaptive penalization. Our theoretical analysis confirms that the FDoS estimator follows the same results and shows that only the FadDoS estimator fulfills the oracle property under mild conditions requiring an initial estimator. To demonstrate the asymptotic behaviors, the following assumptions are required: 

\noindent{(A.1)} $||X||_{2} \leq c_{1} < \infty$ almost surely for some constant $c_{1}$. 

\noindent{(A.2)} $\beta(t)$ satisfies Hölder condition: for some integer $\xi$ and $\zeta \in [0,1]$, $|\beta^{\xi}(t_{1}) - \beta^{\xi}(t_{2})| \leq c_{2}|t_{1}-t_{2}|^{\zeta}$ for some constant $c_{2}$, and $t_{1}, t_{2} \in \mathcal{T}$. We denote $\delta \overset{\text{def}}{=} \xi + \zeta$ and assume that $5/2 < \delta \leq d$, where $d$ is the degree of B-splines.

Assumptions (A.1) and (A.2) are identical with (H1) and (H3) of \cite{cardot}. (A.2) requires the coefficient function $\beta(t)$ to be sufficiently smooth and have $\delta$ derivatives. 

Recall that $\lambda_{1}$, $\lambda_{2}$, and $\varphi$ are tuning parameters for the proposed estimators. For simplicity of the conditions, we choose $\varphi = \lambda_{2}^{2}$ to reduce the number of tuning parameters. In addition, we assume the following condition of choosing values of $M_{n}$.

\noindent{(A.3)} $M_{n} \asymp n^{1/(2\delta-1)}$, which is defined such that both $M_{n}/n^{1/(2\delta-1)}$ and $n^{1/(2\delta-1)}/M_{n}$ are bounded.   

We also provide some notations for illustrating the following properties. Coefficient function $\beta_{l}(t)$ is divided into two regions: $I_{1}(\beta_{l}) = \{I \subset \mathcal{T}:  \sup |\beta_{l}(t)| > D_{l}M_{n}^{-\delta},  \text{for all} \ t \in I\}$ for some positive constant $D_{l}$, and the complementary part is $I_{0}(\beta_{l}) = \{I \subset \mathcal{T}: 0 \leq \sup |\beta_{l}(t)| \leq D_{l}M_{n}^{-\delta},  \text{for all} \ t \in I\}$. It is clear that the zero subregion $N_{0}(\beta_{l})$ is the subset of $I_{0}(\beta_{l})$ for sufficient large $D_{l}$. We further define $\mathcal{B}$ as the set consisting of $k$ basis coefficients whose corresponding basis functions $B_{k}(t)$ have support inside $I_{1}(\beta_{l})$, i.e., $\{k \in \{1,\dots,M_{n}+d\}: b_{k} \neq 0\}$. Let $\mathcal{A} = \{l=\{1, \dots, J\}: \sup|\beta_{l}(t)| > D_{l}M_{n}^{-\delta}, \text{for all} \ t \in \mathcal{T}\}$ be the group of nonzero coefficient functions. 

The first theoretical result demonstrates that our proposed estimator converges at an optimal rate in approximating the true function. It is proved through the estimation error between the proposed estimator and B-spline approximant $\beta_{l}^{\alpha}(t)$ from the family of $\mathcal{S}_{l}$ with a corresponding vector of basis coefficient $\boldsymbol{\alpha}_{l}$, while $\beta_{l}^{\alpha}(t)$ converges to the true function as stated in the following lemma, which is the same as Theorem XII(6) in \cite{deboor}. 
\begin{lemma}
There exists $\beta^{\alpha}(t)$ such that $||\beta - \beta^{\alpha}||_{\infty} \leq C_{1}M_{n}^{-\delta}$ for some constant $C_{1} \geq 0$, where $\beta^{\alpha}(t) \overset{\text{def}}{=} \boldsymbol{B}^{T}(t)\boldsymbol{\alpha}$.
\end{lemma}
It is further noted that the B-spline approximant can preserve the sparsity of the true function because the basis coefficients can be equal to zero if the support of the corresponding basis function falls in the zero subregion $N_{0}(\beta_{l})$. Thus, the asymptotic behavior of the proposed estimator is stated in the following theorem. 

\begin{theorem}[Convergence]
 Under {\normalfont(A.1)-(A.3)}, the penalized estimator satistifies
\begin{equation*}
      ||\hat{\beta}_{l} - \beta_{l}||_{\infty} = O_{p}(M_{n}n^{-1/2}), \ \ l=1,\dots,J,
 \end{equation*}
if $\lambda_{1} = O(M_{n}^{1/2}n^{-1/2})$ and $\lambda_{2}^{2} = O(n^{-1/2})$ for {\normalfont FDoS}; or if $\lambda_{1}\phi_{1} = O_{p}(M_{n}^{1/2}n^{-1/2})$ and $\lambda_{2}^{2}\phi_{2} = O_{p}(n^{-1/2})$ for {\normalfont FadDoS}, where $\phi_{1} = \sup_{l \in \mathcal{A}}||\check{\beta}_{l}||_{1}^{-a}$ and $\phi_{2} = \sup_{l \in \mathcal{A}}||\check{\beta}_{l}||_{2}^{-a}$ are stochastic quantities.
\end{theorem}

The above theorem highlights that there always exists a local solution $\hat{\beta}_{l}(t)$ for the $l$th functional covariate with probability approaching one, while the convergence rate is bounded by $O_{p}(M_{n}n^{-1/2})$.

The oracle property of the functional linear model with multiple functional covariates can be stated under the double-sparsity structure. For global sparsity, the estimator can identify all true nonzero coefficient functions with probability approaching one. For local sparsity, the estimator further recovers zero subregions with probability approaching one and is asymptotically pointwise normally distributed. 

The following theorem is developed for the asymptotic distribution of the FDoS estimator. The proof is through finding the limiting distribution of the vector of B-spline basis coefficient as $n$ goes to infinity, which is also the minimizer of a convex function $V^{(l)}_{1}$. Additionally, we assume a positive-definite matrix, $\boldsymbol{C}_{l} = (M_{n}/n)\boldsymbol{U}_{l}^{T}\boldsymbol{U}_{l}$, whose minimum and maximum eigenvalues are bounded away from zero and infinity with probability approaching one, see Lemma 2 in the supplementary for more details.  
\begin{theorem}
Under {\normalfont (A.1)-(A.3)}, if $\lambda_{1}(n/M_{n})^{1/2} \rightarrow \gamma_{1} \geq 0$ and $\lambda_{2}^{2}n^{1/2} \rightarrow \gamma_{2} \geq 0$,
\begin{equation*}
  (n/M_{n})^{1/2}(\hat\beta_{l}(t) - \beta_{l}(t)) \overset{d}{\rightarrow}  \boldsymbol{B}^{T}_{l}(t)\boldsymbol{v}^{*}_{l}, \ \ l=1,\dots,J,
\end{equation*}
where $\boldsymbol{v}^{*}_{l} = \argmin_{\boldsymbol{v} \in \mathbb{R}^{M_{n}+d}} V^{(l)}_{1}(\boldsymbol{v})$ such that
\begin{equation*}
V_{1}^{(l)}(\boldsymbol{v}) = \boldsymbol{v}^{T}\boldsymbol{C}_{l}\boldsymbol{v} - 2\boldsymbol{W}_{l}^{T}\boldsymbol{v} + \gamma_{1}\Gamma_{1}^{(l)}(\boldsymbol{v}) + \gamma_{2}\Gamma_{2}^{(l)}(\boldsymbol{v}),    
\end{equation*}
and some random vector $\boldsymbol{W}_{l} \sim N(\boldsymbol{0}, \sigma^{2} \boldsymbol{C}_{l})$, $\Gamma_{1}^{(l)}(\boldsymbol{v}) = \sum_{k=1}^{M_{n}+d}\big\{
    |v_{k}|\mathbb{I}(\alpha_{lk} = 0) + v_{k}\text{sgn}\big(\alpha_{lk}\big) \mathbb{I}(\alpha_{lk} \neq 0) \big\}$ and $\Gamma_{2}^{(l)}(\boldsymbol{v}) = ||\boldsymbol{v}||_{2}\mathbb{I}(\boldsymbol{\alpha}_{l} = \boldsymbol{0}) + (\boldsymbol{v}^{T}\boldsymbol{\alpha}_{l}/||\boldsymbol{\alpha}_{l}||_{2}) \mathbb{I}(\boldsymbol{\alpha}_{l} \neq \boldsymbol{0})$.
\end{theorem}
In Thoerem 2, $\Gamma_{1}^{(l)}$
and $\Gamma_{2}^{(l)}$ are brought by estimation error with respect to functional $\ell_{1}$ and $\ell_{1,2}$ respectively. The above theorems imply that when $\lambda_{1} = O(M_{n}^{1/2}n^{-1/2})$ and $\lambda_{2} = O(n^{-1/4})$, the FDoS estimate is $\sqrt{n}/M_{n}$-consistent, but neither global nor local sparsity structures can be recovered with high probability, as stated in the following Proposition 1 under the assumption that $\gamma_{1} < \infty$ and $\gamma_{2} < \infty$.


\begin{proposition}Under {\normalfont(A.1)-(A.3)}, if $\lambda_{1}(n/M_{n})^{1/2} \rightarrow \gamma_{1} \geq 0$ and $\lambda_{2}^{2}n^{1/2} \rightarrow \gamma_{2} \geq 0$, the {\normalfont FDoS} estimator satisfies the following:

\normalfont{1.} $\lim_{n \rightarrow \infty} P(\hat{\mathcal{A}}_{n}=\mathcal{A}) < 1$;

\normalfont{2.} \textit{For $t \in I_{0}(\beta_{l})$ such that $l \in \mathcal{A}$, $P(\hat{\beta}_{l}(t)=0) < 1$.}
\vspace{0.2cm}   
\end{proposition}  

Proposition 1 shows that the FDoS estimator fails to enjoy the oracle property. An appropriate adaptive penalization discussed in Section 2.3 and termed as FadDoS is a promising way to address this issue. The below theorem shows that the FadDoS estimator enjoys the oracle property if we can ensure that adaptive weights satisfy mild conditions. For example, the initial estimator is chosen to be the smoothing spline estimator proposed by \cite{cardot}, which is a $\sqrt{n}/M_{n}$-consistent estimator for $\beta_{l}(t)$. The following results show that the FadDoS estimator achieves the oracle property with proper convergence rates of $\lambda_{1}$ and $\lambda_{2}$.

\begin{theorem}[Oracle Property of FadDoS] 
Under {\normalfont (A.1)-(A.3)}, if $\lambda_{1}(n/M_{n})^{1/2} \rightarrow 0$ and $\lambda_{2}^{2}n^{1/2} \rightarrow 0$; $\lambda_{1}n^{(a+1)/2}/M_{n}^{a+1/2} \rightarrow \infty$ and $\lambda_{2}^{2}n^{(a+1)/2}/M_{n}^{a} \rightarrow \infty$, the {\normalfont FadDoS} estimator satisfies the following:

{\normalfont 1.} {\normalfont (Consistency of global variable selection)} $\lim_{n \rightarrow \infty} P(\hat{\mathcal{A}}_{n} = \mathcal{A})=1$. 

\ \ \ {\normalfont (Consistency of local zero-subregion identification)} For $t \in I_{0}(\beta_{l})$ such that $l \in \mathcal{A}$, $\lim_{n \rightarrow \infty} P(\hat{\beta}_{l}(t)=0)=1$.

{\normalfont 2.} For $t \in I_{1}(\beta_{l})$ such that $l \in \mathcal{A}$, we have $(n/M_{n})^{1/2}(\hat{\beta}_{l}(t) - \beta_{l}(t)) \overset{d}{\rightarrow} N(0, \Sigma_{lt})$, where $\Sigma_{lt} = \sigma^{2}\boldsymbol{B}_{l}^{T}(t)(\boldsymbol{C}_{l})_{\mathcal{B}\mathcal{B}}^{-1}\boldsymbol{B}_{l}(t)$ is the covariance matrix knowing the true subset model, where $(\boldsymbol{C}_{l})_{\mathcal{B}\mathcal{B}}$ be the sub-matrix only with column indices in $\mathcal{B}$. 
\end{theorem}

\section{Simulation}
\label{sec:simu}

We conduct extensive simulation studies to illustrate the global sparsity and local sparsity of the proposed method and compare it with existing methods to evaluate the model performance. 

The true functional linear model is $Y_{i}  = \sum_{j=1}^{10}\int_{0}^{1}X_{ij}(t)\beta_{j}(t)dt + \epsilon_{i}, \ \ i=1,\dots,n$, where $\epsilon_{i} \sim N(0,\sigma_{\epsilon}^{2})$ The measurement error $\sigma_{\epsilon}$ is chosen so that signal-to-noise ratio equals to 4. We generate functional covariates $X_{ij}(t)$ by letting $X_{ij}(t) = \sum_{k}a_{ijk}B_{jk}(t), t \in [0,1]$, where $a_{ijk} \sim N(0,1)$ and $B_{jk}(t)$ is a B-spline basis function defined by order 4 and 50 equally spaced knots. Three different types of coefficient functions are considered, each representing a unique condition: \\
(i) $\beta_{1}(t)$ has a zero subregion 
$\beta_{1}(t) = 2\sin{(3\pi t)}$ for $ 0 \leq t \leq 1/3$; $\beta_{1}(t) = 0$ for $1/3 < t < 2/3$; $\beta_{1}(t) = -2\sin{(3\pi t)}$, for $2/3 \leq t \leq 1$; \\
(ii) $\beta_{2}(t)$ has no zero subregion but two crossings at zero, such that $\beta_{2}(t) = 1.5t^{2} + 2\sin{(3\pi t)}$; \\
(iii) $\beta_{j}(t)=0$ for $j=3,\dots,10$, indicating that the functional covariate has no contribution to the response throughout the entire time domain. We independently simulate 100 replicates. Some examples of generated data are provided in the supplementary file.

The performance of the estimator is evaluated from two aspects reflecting the realization of global sparsity and local sparsity respectively. The global sparsity, which is equivalent to functional variable selection, can be assessed by the true positive rate (TPR) and the true negative rate (TNR) respectively presenting the proportion of significant or insignificant functional covariates accurately retained or excluded from the model. For local sparsity, we use the integrated squared error (ISE) to measure the difference between the estimate and the underlying truth, defined as  $\text{ISE}(\hat{\beta}_{j}) = (1/|\mathcal{T}|)\int_{\mathcal{T}}\big(\hat{\beta}_{j}(t) - \beta_{j}(t)\big)^{2}dt$. It can be further decomposed into $\text{ISE}_{0}(\hat{\beta}_{j})$ and $\text{ISE}_{1}(\hat{\beta}_{j})$ for local zero subregion $N_{0}(\beta_{j})$ and nonzero subregion $N_{1}(\beta_{j})$. The local sparsity can be evaluated via $\text{ISE}_{0}$. Additionally, the predictive performance measured by the prediction mean squared error (PMSE) on the testing data set is also critical to examine the effectiveness of the estimator.

\subsection{Tuning Parameter Effects}

\begin{figure*}[t!]
    \centering
    \includegraphics[width=0.825\textwidth, height=0.55\textwidth]{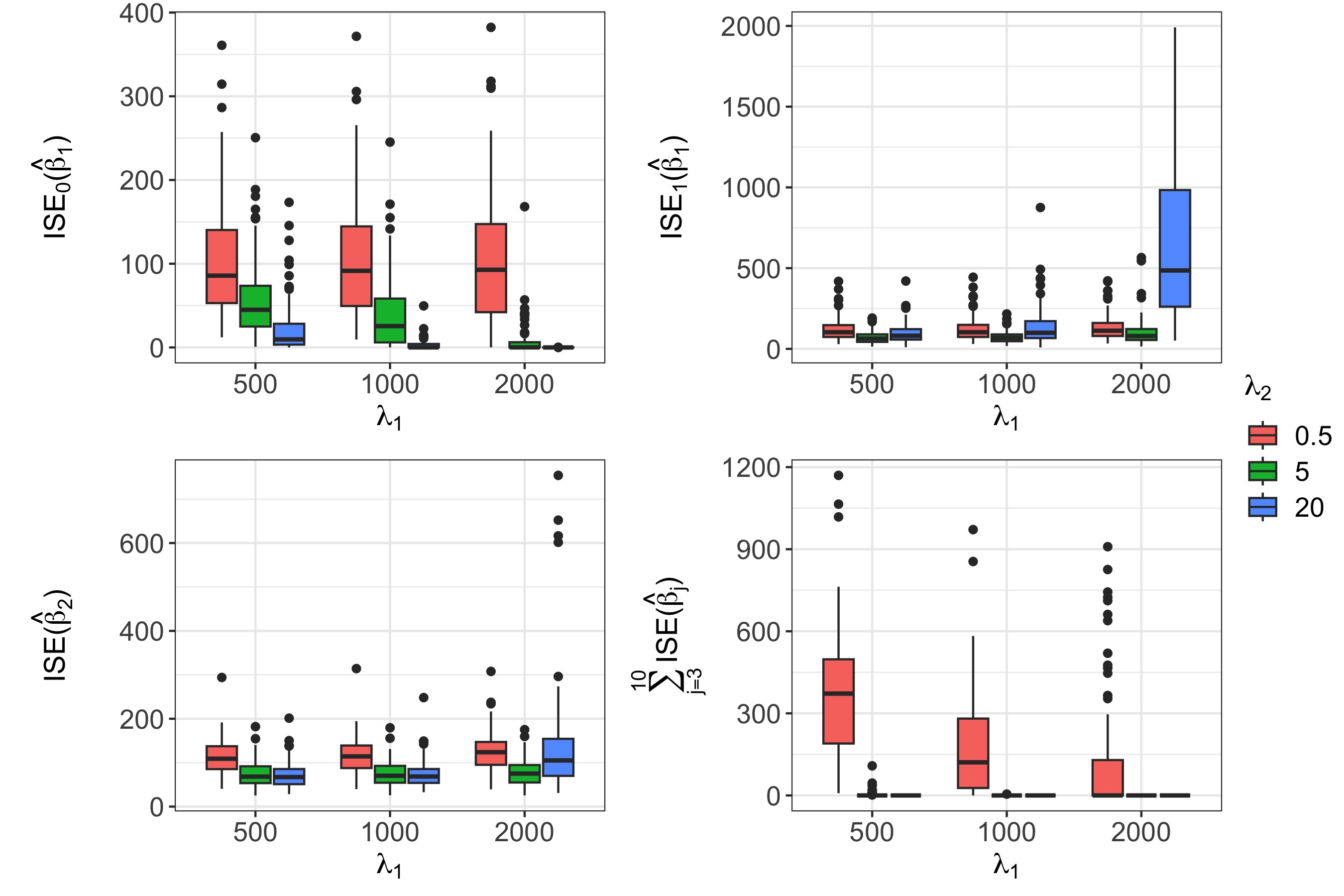}
    \caption{The ISE performance of the FadDoS estimator with varying $\lambda_{1}$ and $\lambda_{2}$, $\varphi =$ 1e-5 over 100 simulation replicates. The training sample size is 200.}
    \label{fig:varying_paras}
\end{figure*}

Figure \ref{fig:varying_paras} shows the effect of tuning parameters $\lambda_{1}$ and $\lambda_{2}$ on the behavior of the FadDoS estimator by controlling the strength of penalization. Increasing $\lambda_{1}$ and $\lambda_{2}$ shrink $\hat{\beta}_{1}(t)$ in zero subregion and all $\hat{\beta}_{j}(t)$, $j=3,\dots,10$ towards zero, while the procedure delivers exactly the opposite results regarding nonzero subregions of $\hat{\beta}_{1}(t)$and $\hat{\beta}_{2}(t)$. Thus, they provide an elastic-net-like estimation balancing zero and nonzero regions of coefficient functions. Table \ref{table: varying_paras} further provides quantitative evaluation of tuning parameter effects. The minimization of PMSE is achieved under the trade-off between local and global sparsity parameters. Moreover, the global sparsity parameter $\lambda_{2}$ primarily administers the overall sparseness of the functional estimates, thereby achieving functional variable selection by successfully identifying all zero coefficient functions. Further discussion of the effect and choice of $\varphi$ can be found in the supplementary materials.

\begin{table}[t!]
    \centering
    \scalebox{0.85}{
\begin{tabular}{c c c c c}
\hline
 & $\lambda_{1}$ & \multicolumn{3}{c}{$\lambda_{2}$} \\[0.5ex]
 \cline{3-5}
 &  &  0.5 & 5 & 20 \\[0.5ex]
 \hline
\multirow{3}{*}{$\text{PMSE} (\times 10^{-2})$} & 500 & 3.29 (0.50) & 2.51 (0.18) & 2.52 (0.18) \\[0.5ex]
 & 1000  & 2.95 (0.42) & 2.49 (0.18) & 2.56 (0.22) \\[0.5ex]
 & 2000 & 2.85 (0.42) & 2.51 (0.21) & 3.41 (0.87) \\[0.5ex]
 \hline 
 \multirow{3}{*}{$\text{avgTPR}$} & 500 & 1.00 (0.00) & 1.00 (0.00) & 1.00 (0.00) \\[0.5ex]
 & 1000  & 1.00 (0.00) & 1.00 (0.00) & 1.00 (0.00)  \\[0.5ex]
 & 2000 & 1.00 (0.00)  & 1.00 (0.00) & 0.96 (0.14) \\[0.5ex]
 \hline
 \multirow{3}{*}{$\text{avgTNR}$} & 500 & 0.33 (0.21) & 0.98 (0.05) & 1.00 (0.00) \\[0.5ex]
 & 1000  & 0.67 (0.27) & 1.00 (0.01) & 1.00 (0.00) \\[0.5ex]
 & 2000 & 0.84 (0.24) & 1.00 (0.00) & 1.00 (0.00) \\[0.5ex]
 \hline
        \end{tabular}}
        \caption{The PMSE ($\times 10^{-2}$), average TPR (avgTNR), and average TNR (avgTNR) performance of the FadDoS estimator with varying $\lambda_{1}$ and $\lambda_{2}$, $\varphi = $1e-5.
        Each entry is the average of 1000 test examples throughout 100 simulation replicates. The entry in the parenthesis corresponds to the standard deviation. The training sample size is 200 and the test sample size is 1000.}
        \label{table: varying_paras}
\end{table}

\subsection{Comparison with Existing Methods} 

We compare our proposed estimators with the aforementioned estimation methods fGL (\citet{gertheiss}), fL (\citet{mingotti}), and SLoS (\citet{lin}). Tuning parameters for all estimators are selected by five-fold cross-validation to ensure optimal performance of each estimator. Although the software of fL can not be identified, we can still implement it by our algorithm it is a special case of the FDoS estimator when $\lambda_{2} = 0$.

\begin{figure*}[b!]
    \centering
     \includegraphics[width=0.825\textwidth, height=0.55\textwidth]{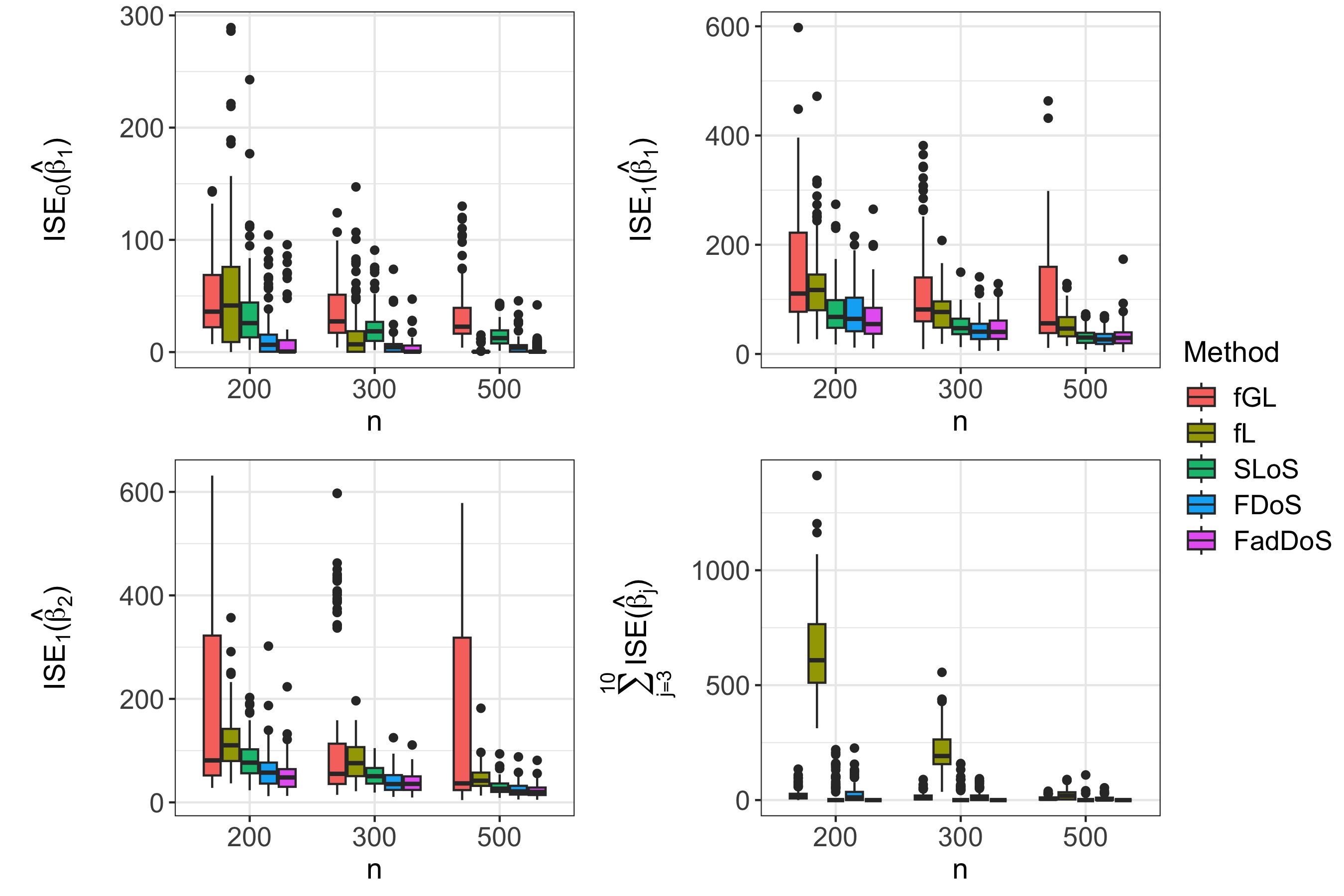}
    \caption{Comparison of ISE performance for fGL (\citet{gertheiss}), fL(\citet{mingotti}), SLoS (\citet{lin}), FDoS, and FadDoS. }
    \label{fig:comparison}
\end{figure*}

\begin{table}[thb!]
    \centering  \hspace*{-12mm}
\scalebox{0.8}{
\begin{tabular}{|c c c c c|} 
\hline
 & Methods & PMSE($\times 10^{-2}$) & avgTPR & avgTNR \\[0.5ex]
\hline
\multirow{6}{*}{n=200} & fGL & 2.57(0.19) & 1.00(0.00) & 0.45(0.04) \\[0.5ex]
 & fL & 3.76(0.51) & 1.00(0.00) & 0.01(0.01) \\[0.5ex]
 & SLoS & 2.49(0.21) & 1.00(0.00) & 0.96(0.02) \\[0.5ex]
 & FDoS  & 2.52(0.91) & 1.00(0.00) & 0.78(0.04) \\[0.5ex]
 & FadDoS  & 2.44(0.18) & 1.00(0.00) & 1.00(0.00) \\[0.5ex]
 \hline  
 \multirow{6}{*}{n=300} & fGL & 2.45(0.17) & 1.00(0.00) & 0.49(0.04)\\[0.5ex]
 & fL & 2.81(0.25) & 1.00(0.00) & 0.17(0.04) \\[0.5ex]
 & SLoS & 2.40(0.17) & 1.00(0.00) & 0.97(0.02) \\[0.5ex]
 & FDoS & 2.41(0.17) & 1.00(0.00) & 0.81(0.04) \\[0.5ex]
 & FadDoS & 2.38(0.16) & 1.00(0.00) & 1.00(0.00) \\[0.5ex]
 \hline
 \multirow{6}{*}{n=500} & fGL & 2.36(0.15) & 1.00(0.00) & 0.50(0.06)\\[0.5ex]
 & fL & 2.38(0.16) & 1.00(0.00) & 0.77(0.03) \\[0.5ex]
 & SLoS & 2.31(0.15) & 1.00(0.00) & 0.99(0.01) \\[0.5ex]
 & FDoS & 2.32(0.15) & 1.00(0.00) & 0.86(0.04) \\[0.5ex]
 & FadDoS & 2.31(0.15) & 1.00(0.00) & 1.00(0.00) \\[0.5ex]
 \hline
        \end{tabular}}
        \caption{PMSE($\times 10^{-2}$), average TPR (avgTPR), and average TNR (avgTNR) of the proposed estimators FDoS and FadDoS as well as fGL (\citet{gertheiss}), fL(\citet{mingotti}), and SLoS (\citet{lin}). The test sample size is 1000. The entry in the parenthesis corresponds to the standard deviation among 100 simulation replicates.} 
        \label{table:comparison}
\end{table}

Figure \ref{fig:comparison} shows the ISE performance of all estimators for different types of coefficient functions, while Table \ref{table:comparison} summarizes PMSE and variable selection accuracy. The proposed FDoS and FadDoS demonstrate comparable results to SLoS, and outperform fGL and fL overall. This highlights the value of modeling both global and local sparsity simultaneously, which fGL and fL fail to achieve, limiting their ability to handle such complex coefficient functions. In particular, utilizing only functional $\ell_{1,2}$ (fGL) or functional $\ell_{1}$ (fL) penalties results in either under- or over-shrinkage.

More importantly, the use of proper adaptive weights in the FadDoS estimator can lead to more accurate estimation of zero and nonzero regions of $\beta_{1}(t)$ and $\beta_{2}(t)$ simultaneously, and at the same time reduce the number of false negatives, namely avoiding nonzero estimates of true zero functions, as suggested by the higher average TNR for the eight true zero coefficients. It is noteworthy that FadDoS slightly outperforms SLoS. One explanation is that the penalization terms in FadDoS are weighted by the initial coefficient function estimates to control both global and local sparsity, while the sparsity parameters for every functional covariate designed in SLoS are identical. Therefore, the SLoS estimator might not regularize both global and local shrinkage simultaneously. Additionally, SLoS demonstrates worse estimation at the last coefficient function. While tail fluctuations when using polynomial splines are expected, FadDoS alleviates this wiggliness issue by embedding the roughness penalty within the functional $\ell_{1,2}$ penalty term rather than placing it independently as SLoS. 

\section{Application}
\label{sec:app}

In this section, the proposed FadDoS is applied to the motivating Kinect study to examine the association between elderly mobility assessments and their multi-joint movements. A total of 50 subjects aged between 68 and 89 years from four Neighbourhood Elderly Centres (NECs) in Hong Kong participated in the study. The participants performed a functional balance test called Timed Up and Go (TUG) test (\citet{podsiadlo}), during which Kinect sensor monitored the three-dimensional displacements of 25 skeletal joints. The whole process of TUG test includes standing up from the chair, walking forward, turning around, walking back to the chair, and sitting down, as illustrated in Figure S2. We use their Balance Evaluation Systems Test (BESTest) scores as the measure of general functional balance and walking capability (\citet{horak}). In the pre-processing step, since the shorter completion time itself reflects better mobility (\citet{nielsen}), we padded the shorter signals till all observations have an equal time length to preserve a series of sequential raw actions, instead of warping time to align all movement signals. Additionally, all displacement signals are interpolated to ensure the same sampling rate. We calculate velocity by taking the first derivative of displacement signals. It is prioritized over displacement because they are less likely to be affected by different positioning of the Kinect sensor. In addition, previous research has shown interesting frequency patterns during walking (\citet{kestenbaum, morgan}), and therefore we include frequency domain knowledge to provide a more insightful understanding of joint movement patterns for the entire activity. The frequency domain information is obtained via Fourier transform. In scalar-on-function regression model, we regress the average BESTtest score on velocity and frequency of three dimensions respectively in order to identify associated joints and investigate in detail how joint movements are associated with mobility assessments over the time and frequency domains.

\begin{figure}[!b]
\centering
\begin{subfigure}{.3\textwidth}
  \centering \hspace*{-20mm}
  \includegraphics[width=1\textwidth]{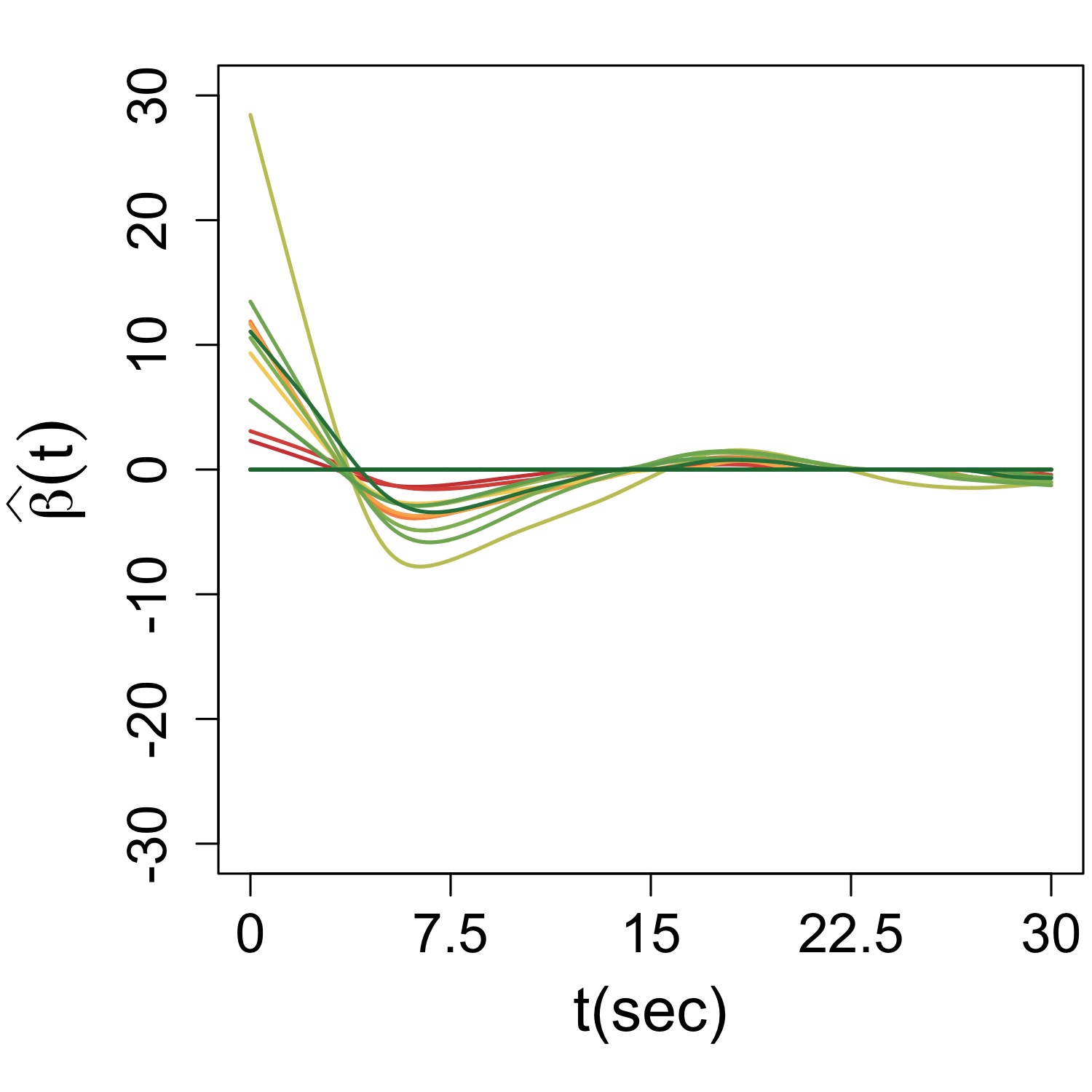}
  \caption{\hspace*{8mm}}
\end{subfigure}
\begin{subfigure}{.3\textwidth}
  \centering \hspace*{-20mm}
  \includegraphics[width=1\textwidth]{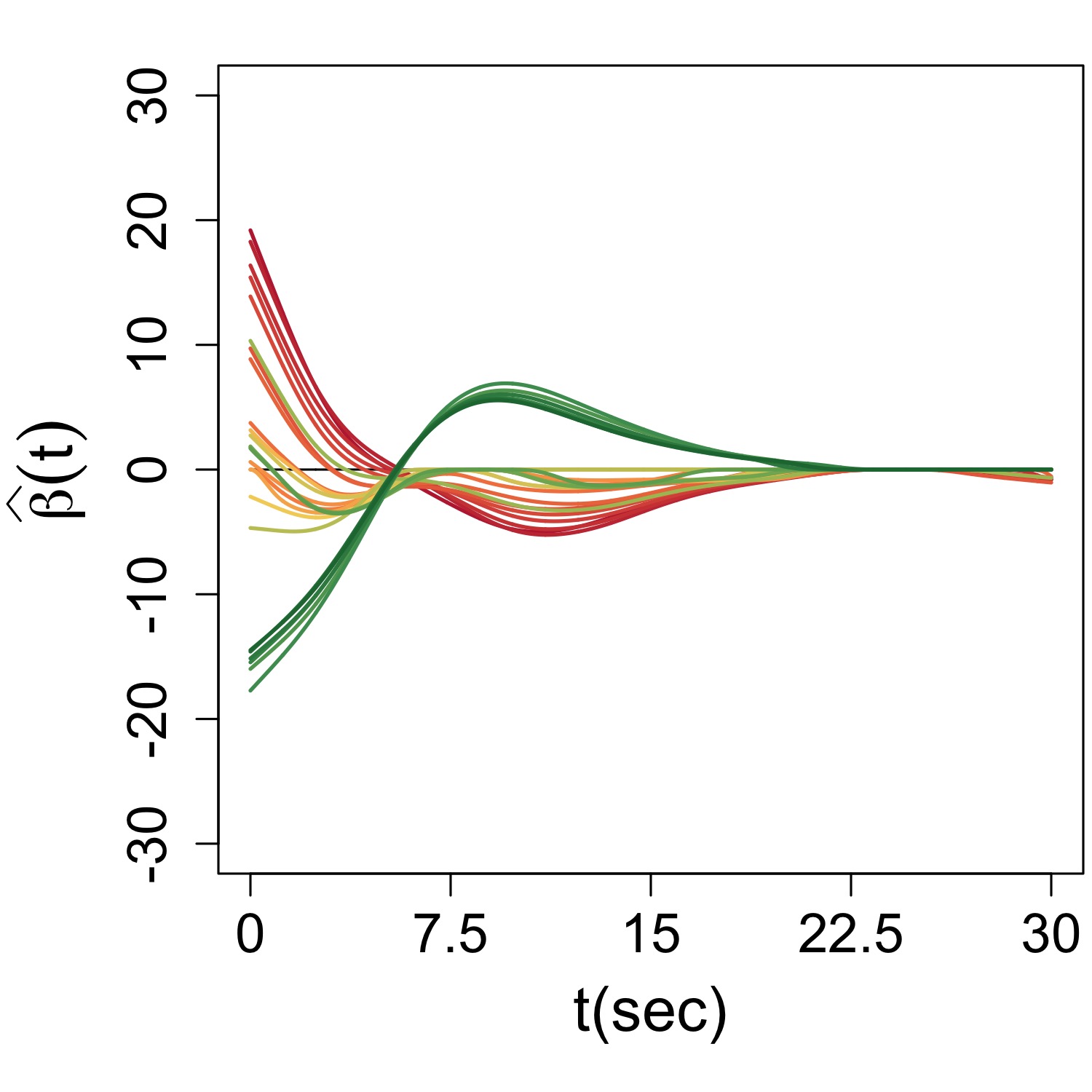}  
  \caption{\hspace*{7mm}}
\end{subfigure}
\begin{subfigure}{.3\textwidth}
  \centering \hspace*{-10mm}
  \includegraphics[width=1.4\textwidth, height=1\textwidth]{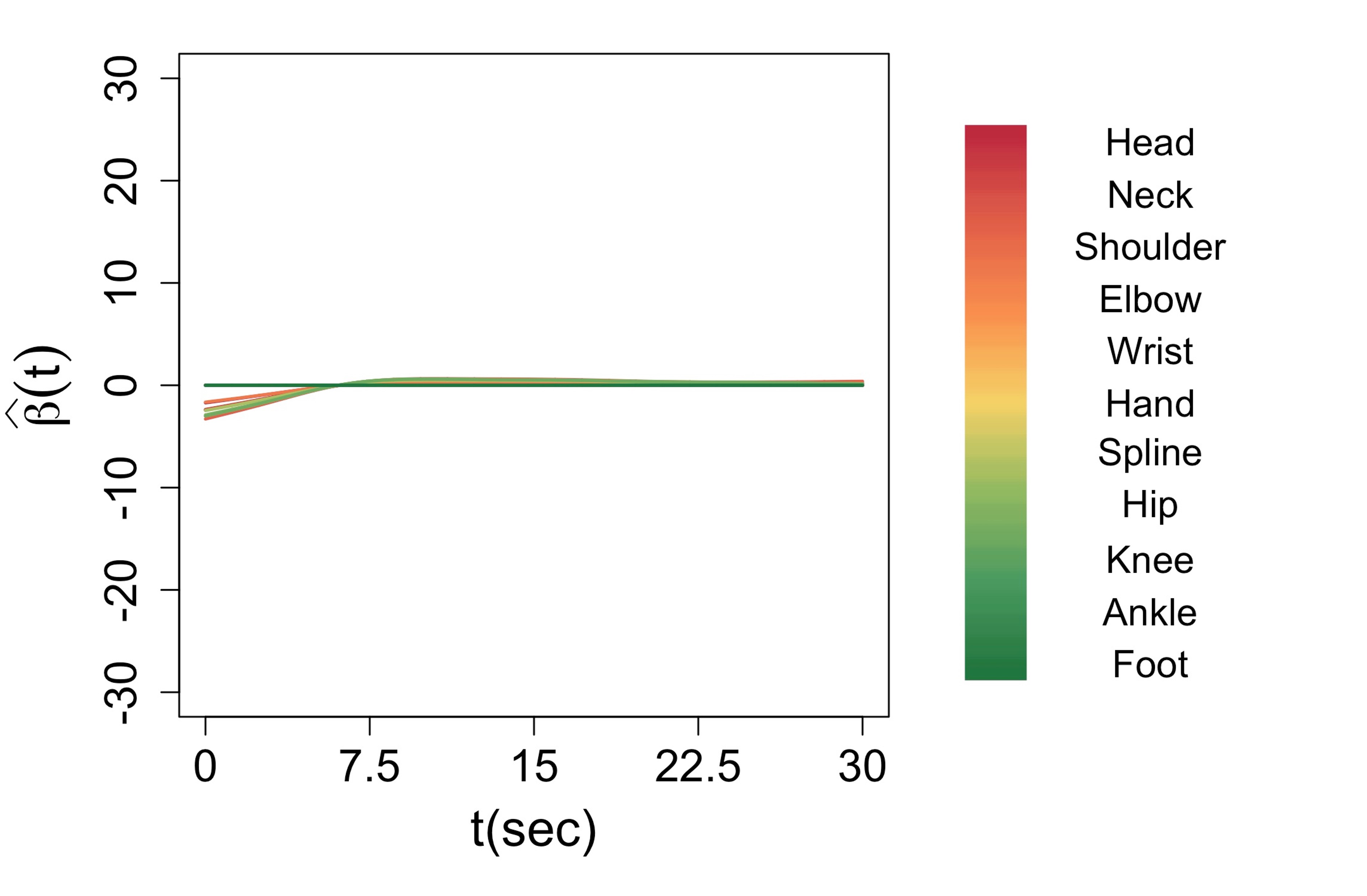}  
  \caption{\hspace*{10mm}}
\end{subfigure}
\caption{The estimated coefficient functions of (a) horizontal velocity, (b) vertical velocity, and (c) walking velocity. Each curve represents the estimated coefficient function of a joint and is color coded as Figure \ref{fig:setup}(b).}
\label{fig:realvel}
\end{figure}

Figure \ref{fig:realvel}(a) shows that the horizontal velocity of all joints shares an almost identical pace and therefore yields similar estimates. The estimates are positive till around 3 seconds when the healthy elderly stands up, which implies that faster ascending comes with good mobility. All estimates are close to zero around 18 seconds when the majority of the subjects completed the test. Figure \ref{fig:realvel}(b) exhibits three clusters of behavior patterns related to joints in head, trunk and upper limbs, and lower limbs. The positive coefficients for the upward velocity from head to shoulder in the beginning also reveal that the swifter standing up the better mobility. Due to the setup of the sensor shown in Figure \ref{fig:setup}(a), the trunk and upper limbs are closer to the position of the Kinect camera, namely the origin of the three-dimensional coordinates, leading to relatively weaker effects. For lower limbs, high velocity during ascending and walking forward implies good physical capability. Moreover, the direction of vertical velocity is reversed when walking backward. It is reflected in the change of the signs in the estimates, all of which cross zero around 7 seconds as the healthy elderly turn. It is observed that the amplitude of the estimated coefficient is maximized at 12 seconds for head or 10 seconds for lower limbs when returning to the chair. The lag is due to the downward motion of head during sitting down. The nonzero estimates of most joints depicted in Figure \ref{fig:realvel}(c) display consistent patterns. Since the stride velocity during walking forward is negative due to the Kinect camera setup, negative estimates before 7 seconds also indicate the faster pace the better mobility assessment.

\begin{figure}[!b]
\centering
\begin{subfigure}{.3\textwidth}
  \centering \hspace*{-20mm}
  \includegraphics[width=1\textwidth]{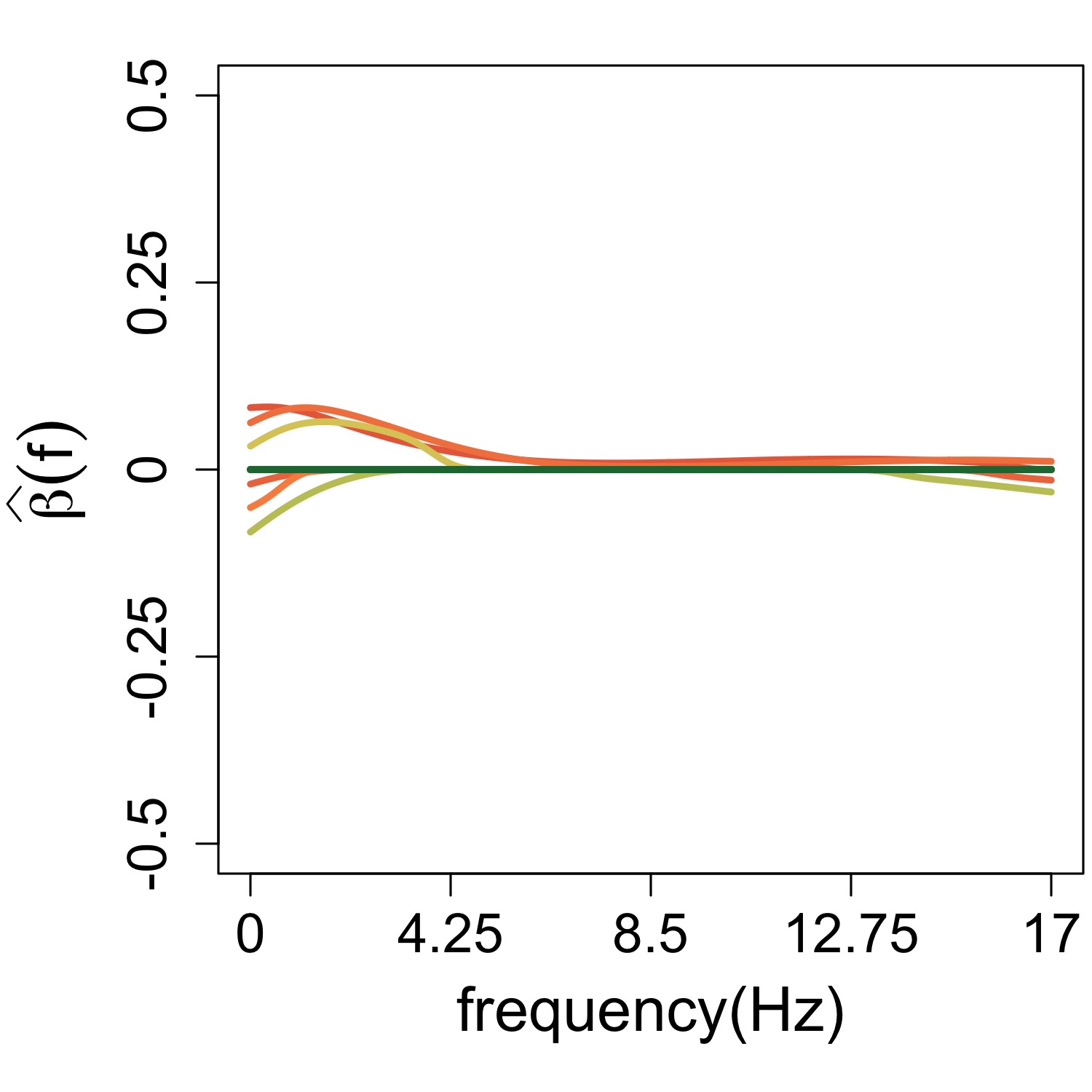}
  \caption{\hspace*{10mm}}
\end{subfigure}
\begin{subfigure}{.3\textwidth}
  \centering  \hspace*{-20mm}
  \includegraphics[width=1\textwidth]{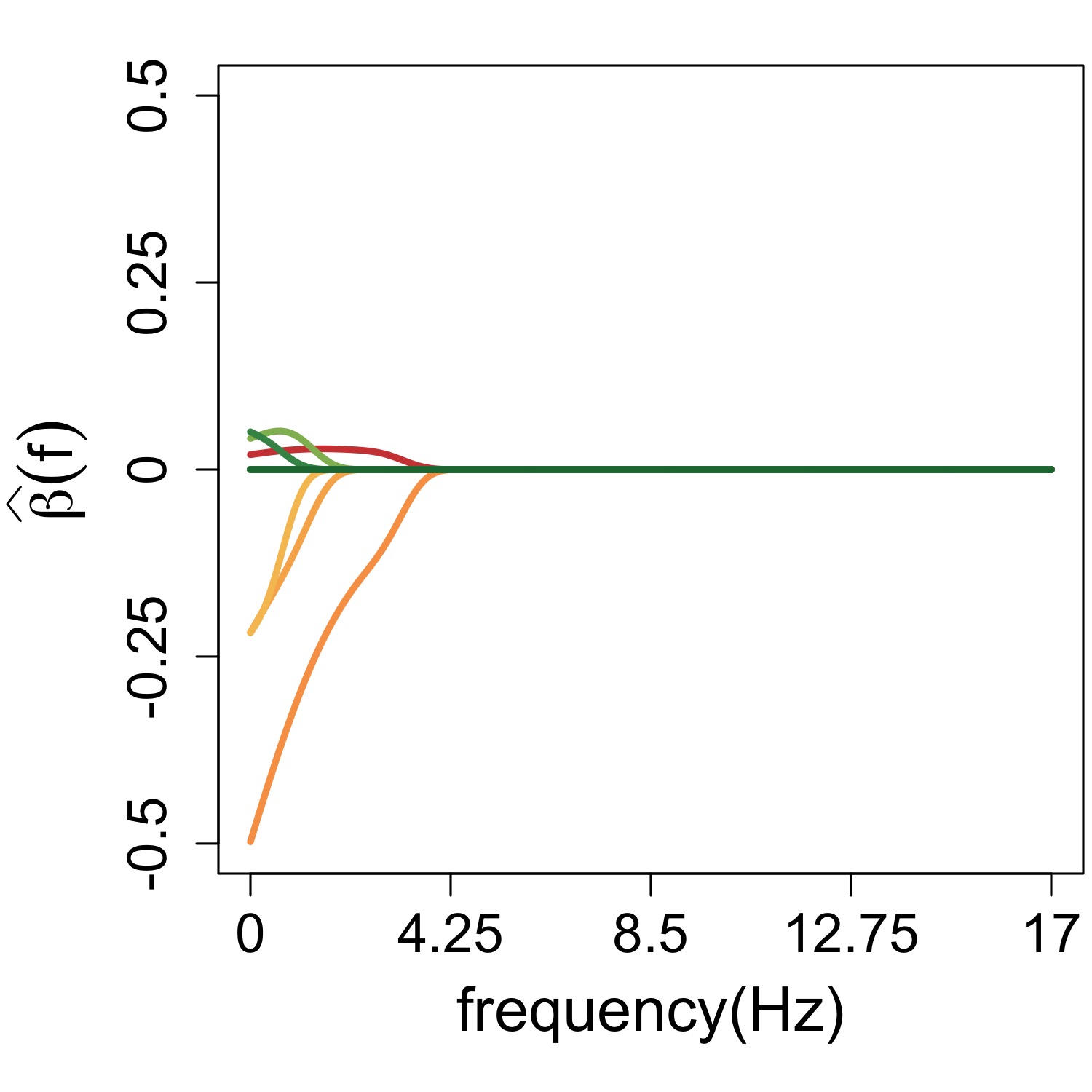}  
  \caption{\hspace*{5mm}}
\end{subfigure}
\begin{subfigure}{.3\textwidth}
  \centering  \hspace*{-10mm}
  \includegraphics[width=1.4\textwidth, height=1\textwidth]{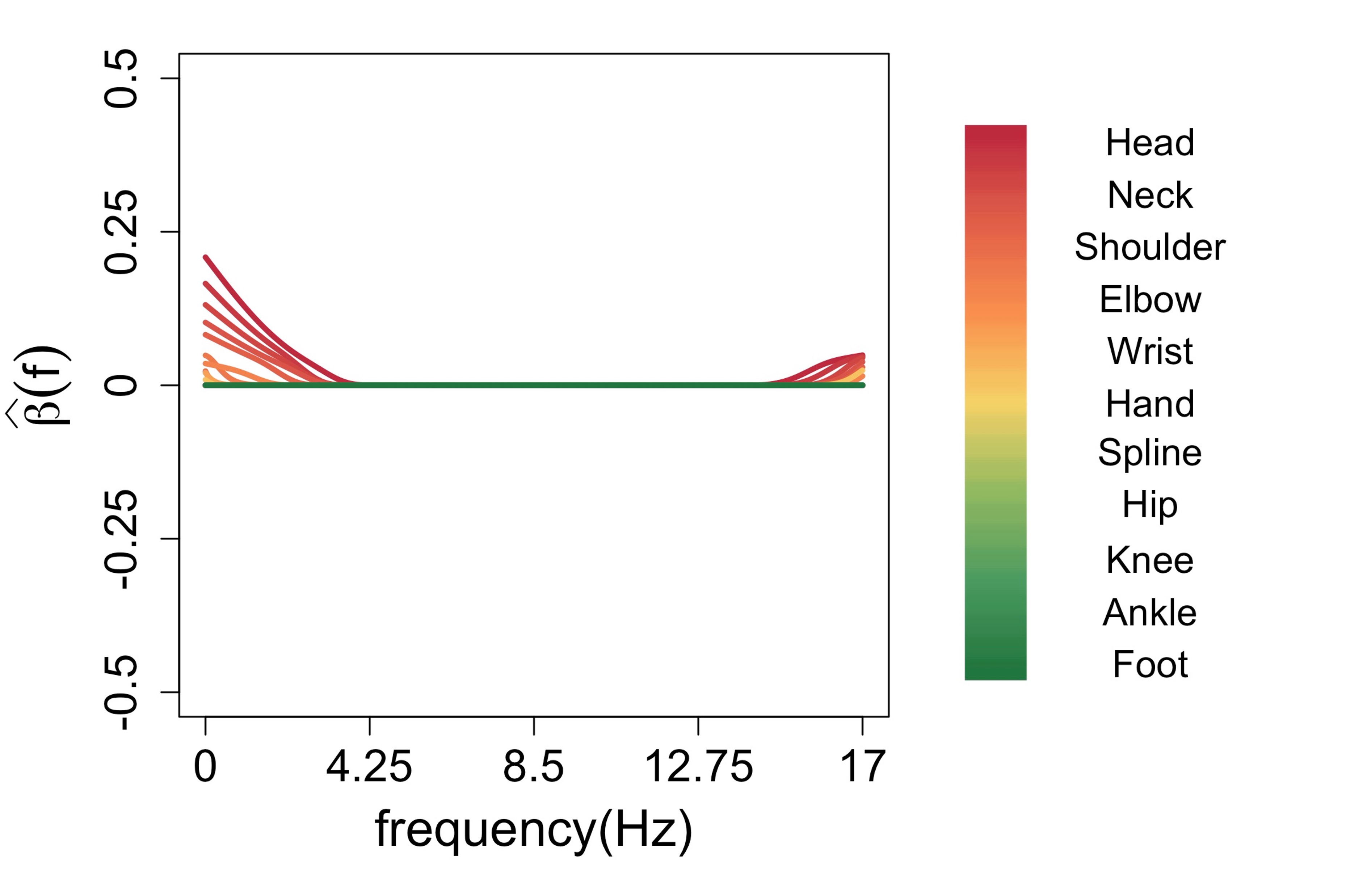}  
  \caption{\hspace*{10mm}}
\end{subfigure}
\caption{The estimated coefficient functions of (a) horizontal frequency, (b) vertical frequency, and (c) walking frequency. Each curve represents the estimated coefficient function of a joint and is color coded as Figure \ref{fig:setup}(b).}
\label{fig:realfreq}
\end{figure}

Figure \ref{fig:realfreq} shows the estimated coefficient functions in the frequency spectrum from 0 Hz to 17 Hz along three directions. Figure \ref{fig:realfreq}(a) demonstrates that only joints of shoulder, elbows, and hands have substantial nonzero effects in part of the horizontal frequency domain, where the low-frequency oscillation reflects the arm's horizontal swinging motion during walking, while the high-frequency oscillation represents the horizontally involuntary and rhythmic shaking probably caused by some disorders and thus associated with lower mobility (\citet{hess}). The estimates of vertical oscillations of the shoulder, spine, and ankles are positive only at low frequencies as shown in Figure \ref{fig:realfreq}(b), representing the natural up-and-down oscillation in the normal stride. However, the negative estimates imply that the vertical swinging movement of hands is generally seen among the low-mobility elderly who spend excessive energy vertically to assist with walking. Additionally, Figure \ref{fig:realfreq}(c) shows that only joints between the head and hands in the walking direction are associated with mobility. The nonzero estimates at the low-frequency domain confirm that the average stride frequency is 2 Hz (\citet{henriksen}). The magnitude of the effect decreases from the top joints to the bottom is likely because intensive head motions tilting forward are compensatory for trunk pitch to stabilize the body (\citet{hirasaki, hirasaki1}). Moreover, the positive estimates of high-frequency oscillation are probably due to normal action tremors present in healthy individuals for active movements (\citet{hess}).

\section{Conclusion}
\label{sec:conc}

While scalar-on-function regression model is appealing for multi-dimensional sensor data, where a large number of functional covariates are collected and scalar responses are measured, the double-sparsity property is critical for the interpretability of the model. Thus, we propose the novel FadDoS estimator, based on which we develop a nice double-sparsity model. In particular, one can concurrently identify important functional covariates and influential nonzero subregions throughout the entire domain. Our estimator is developed based on the functional generalization of sparse group lasso to control local and global sparsity respectively, as well as smoothness. We further enhance it with adaptation where the penalty can be weighted by preliminary estimates. The FadDoS estimator enjoys the oracle property under mild conditions of the initial estimator and therefore it is theoretically sounder than the nonadaptive version. Simulation studies also show that our estimator outperforms existing methods with optimal performance. 

Our work has provided an effective method for analyzing wearable sensor data, especially those that capture high-dimensional functional signals. In the application of Kinect sensor study, we achieved satisfactory performance and identified the associated joints and their detailed association over both time and frequency domains. Furthermore, the FadDoS estimator is generalizable and applicable to other types of sensor devices that collect multi-dimensional signals including acceleration, heart rates, skin impedance, and more, thus promoting the use of advanced sensor devices in health research and having wide applications in the health field.

\bigskip

{\noindent \large\bf Acknowledgments}

\noindent{The authors thank all participants who took part in the study and the researchers who assisted with data collection.}

\bigskip
{\noindent \large\bf Funding}

\noindent{This study was funded by the City University of Hong Kong, Hong Kong SAR, China internal research grant 610473 and 7005892 and the Hong Kong Innovation and Technology Commission (InnoHK Project CIMDA). The funder did not have any influence over the study design, subsequent analysis, or drafting of the manuscript.}

\bibliography{main}

\end{document}



\renewcommand{\baselinestretch}{2}

\markright{ \hbox{\footnotesize\rm Statistica Sinica: Supplement
}\hfill\\[-13pt]
\hbox{\footnotesize\rm
}\hfill }

\markboth{\hfill{\footnotesize\rm FIRSTNAME1 LASTNAME1 AND FIRSTNAME2 LASTNAME2} \hfill}
{\hfill {\footnotesize\rm FILL IN A SHORT RUNNING TITLE} \hfill}

\renewcommand{\thefootnote}{}
$\ $\par \fontsize{12}{14pt plus.8pt minus .6pt}\selectfont


 \centerline{\large\bf  FUNCTIONAL ADAPTIVE DOUBLE-SPARSITY ESTIMATOR}
\vspace{2pt} 
\centerline{\large\bf  FOR FUNCTIONAL LINEAR REGRESSION MODEL WITH}
\vspace{2pt} 
\centerline{\large\bf   MULTIPLE FUNCTIONAL COVARIATES}
\vspace{.4cm} 
\centerline{Cheng Cao\textsuperscript{1}, Jiguo Cao\textsuperscript{2}, Hailiang Wang\textsuperscript{3},  Kwok-Leung Tsui\textsuperscript{4},  Xinyue Li\textsuperscript{1}$^{*}$} 
\vspace{.4cm} 
\centerline{  \textsuperscript{1}\textit{City University of Hong Kong}, \textsuperscript{2}\textit{Simon Fraser University},}
\centerline{  \textsuperscript{3}\textit{The Hong Kong Polytechnic University}, \textsuperscript{4}\textit{Virginia Polytechnic Institute and State University}}
\vspace{.55cm}
 \centerline{\bf Supplementary Material}
\vspace{.55cm}
\fontsize{9}{11.5pt plus.8pt minus .6pt}\selectfont
\let\thefootnote\relax\footnotetext{$^{*}$Corresponding author: Xinyue Li, Email: xinyueli@cityu.edu.hk}

\noindent
This supplementary material includes illustrations of the simulation studies, real data application, and theoretical results of our proposed method. We provide derivations of Alternating Direction Method of Multipliers (ADMM) subproblems in Section S1. The proofs of Theorem 1-Theorem 3 are given in Section S2. The figures to demonstrate simulation performance and Timed Up and Go test are shown in Section S3 and S4. 
\par

\setcounter{section}{0}
\setcounter{equation}{0}
\def\theequation{S\arabic{section}.\arabic{equation}}
\def\thesection{S\arabic{section}}

\fontsize{12}{14pt plus.8pt minus .6pt}\selectfont

\newpage

\section{Derivation of ADMM subproblems}

\subsection{Solution to $\tilde{\boldsymbol{b}}$-update}
Based on the Equation (8), 
\begin{equation*}
\begin{aligned}
    \tilde{\boldsymbol{b}}_{l}^{k+1} &= \argmin_{\tilde{\boldsymbol{b}}_{l} \in \mathbb{R}^{M_{n}+d}} \frac{1}{2}||\boldsymbol{r}_{(-l)}-\widetilde{\boldsymbol{U}}_{l}\tilde{\boldsymbol{b}}_{l}||_{2}^{2} + \lambda_{2} ||\tilde{\boldsymbol{b}}_{l}||_{2}  + (\boldsymbol{u}^{k}_{l})^{T}((\boldsymbol{L}_{l}^{T})^{-1} \tilde{\boldsymbol{b}}_{l}-\boldsymbol{z}^{k}_{l}) + \frac{\rho}{2}||(\boldsymbol{L}_{l}^{T})^{-1} \tilde{\boldsymbol{b}}_{l} - \boldsymbol{z}^{k}_{l}||_{2}^{2} \\
        & = \argmin_{\tilde{\boldsymbol{b}}_{l} \in \mathbb{R}^{M_{n}+d}} \frac{1}{2}||\boldsymbol{r}_{(-l)}-\widetilde{\boldsymbol{U}}_{l}\tilde{\boldsymbol{b}}_{l}||_{2}^{2} + \lambda_{2} ||\tilde{\boldsymbol{b}}_{l}||_{2} +\frac{\rho}{2}||(\boldsymbol{L}_{l}^{T})^{-1} \tilde{\boldsymbol{b}}_{l} - \boldsymbol{z}^{k}_{l} - \boldsymbol{u}^{k}_{l}/\rho||_{2}^{2} \\ 
        & = \argmin_{\tilde{\boldsymbol{b}}_{l} \in \mathbb{R}^{M_{n}+d}} \lambda_{2} ||\tilde{\boldsymbol{b}}_{l}||_{2} + \frac{1}{2}||\hat{\boldsymbol{r}}_{(-l)}^{k} - \widehat{\boldsymbol{U}}_{l}\tilde{\boldsymbol{b}}_{l}||_{2}^{2},
\end{aligned}
\end{equation*}
where $\widehat{\boldsymbol{U}}_{l} = \big(\widetilde{\boldsymbol{U}}_{l}^{T}, \sqrt{\rho}\boldsymbol{L}_{l}^{-1}\big)^{T}$ and $\hat{\boldsymbol{r}}_{(-l)}^{k} = \big(\boldsymbol{r}_{(-l)}^{T}, \sqrt{\rho}(\boldsymbol{z}^{k}_{l} + \boldsymbol{u}^{k}_{l}/\rho\big)^{T})^{T}$. Since $\widehat{\boldsymbol{U}}_{l}$ is not identity, we cannot directly get the solution to the second operator. Fortunately, by following \cite{wangx} we can employ linearization technique to approximate the quadratic term efficiently. We linearize it by replacing  $||\hat{\boldsymbol{r}}_{(-l)}^{k} - \widehat{\boldsymbol{U}}_{l}\tilde{\boldsymbol{b}}_{l}||_{2}^{2}/2$ with
$\big(\widehat{\boldsymbol{U}}_{l}^{T}\big(\widehat{\boldsymbol{U}}_{l}\tilde{\boldsymbol{b}}_{l}^{k} -\hat{\boldsymbol{r}}_{(-l)}^{k} \big)\big)^{T}(\tilde{\boldsymbol{b}}_{l} - \tilde{\boldsymbol{b}}_{l}^{k}) + \frac{\nu_{l}}{2}||\tilde{\boldsymbol{b}}_{l} - \tilde{\boldsymbol{b}}_{l}^{k}||^{2}_{2}$, where the first term is apparently the gradient at $\tilde{\boldsymbol{b}}_{l}^{k}$. Hence,
\begin{equation*}
\begin{aligned}
    \tilde{\boldsymbol{b}}_{l}^{k+1} & = \argmin_{\tilde{\boldsymbol{b}}_{l} \in \mathbb{R}^{M_{n}+d}} \lambda_{2} ||\tilde{\boldsymbol{b}}_{l}||_{2} + \big(\widehat{\boldsymbol{U}}_{l}^{T}\big(\widehat{\boldsymbol{U}}_{l}\tilde{\boldsymbol{b}}_{l}^{k}-\hat{\boldsymbol{r}}_{(-l)}^{k} \big)\big)^{T}(\tilde{\boldsymbol{b}}_{l} - \tilde{\boldsymbol{b}}_{l}^{k}) + \frac{\nu_{l}}{2}||\tilde{\boldsymbol{b}}_{l} - \tilde{\boldsymbol{b}}_{l}^{k}||^{2}_{2} \\
    & = \argmin_{\tilde{\boldsymbol{b}}_{l} \in \mathbb{R}^{M_{n}+d}} \lambda_{2} ||\tilde{\boldsymbol{b}}_{l}||_{2} + \frac{\nu_{l}}{2} ||\tilde{\boldsymbol{b}}_{l} - \tilde{\boldsymbol{b}}_{l}^{k} + \widehat{\boldsymbol{U}}_{l}^{T}\big(\widehat{\boldsymbol{U}}_{l}\tilde{\boldsymbol{b}}_{l}^{k}-\hat{\boldsymbol{r}}_{(-l)}^{k} \big)/\nu_{l}||^{2}_{2} \\
    & = S_{2, \lambda_{2}/\nu_{l}}(\tilde{\boldsymbol{b}}_{l}^{k} - \widehat{\boldsymbol{U}}_{l}^{T}\big(\widehat{\boldsymbol{U}}_{l}\tilde{\boldsymbol{b}}_{l}^{k}-\hat{\boldsymbol{r}}_{(-l)}^{k} \big)/\nu_{l}).
\end{aligned}
\label{eq:b_update}
\end{equation*}
If $\rho$ is fixed, $\tilde{\boldsymbol{b}}_{l}^{k+1}$ will tend to be zero when $\lambda_{2} \geq \nu_{l}||\tilde{\boldsymbol{b}}_{l}^{k} - \widehat{\boldsymbol{U}}_{l}^{T}\big(\widehat{\boldsymbol{U}}_{l}\tilde{\boldsymbol{b}}_{l}^{k}-\hat{\boldsymbol{r}}_{(-l)}^{k} \big)/\nu_{l}||_{2}$.

\subsection{Solution to $\boldsymbol{z}$-update}
Based on the Equation (9), 
\begin{equation*}
\begin{aligned}
        \boldsymbol{z}^{k+1} & = \argmin_{\boldsymbol{z} \in \mathbb{R}^{J\times(M_{n}+d)}} \lambda_{1}||\boldsymbol{z}||_{1} + (\boldsymbol{u}^{k})^{T}(\boldsymbol{D}\tilde{\boldsymbol{b}}^{k+1} - \boldsymbol{z}) + \frac{\rho}{2}||\boldsymbol{D}\tilde{\boldsymbol{b}}^{k+1} - \boldsymbol{z}||_{2}^{2} \\ 
        & = \argmin_{\boldsymbol{z} \in \mathbb{R}^{J\times(M_{n}+d)}} \lambda_{1}||\boldsymbol{z}||_{1} + \frac{\rho}{2}||\boldsymbol{D}\tilde{\boldsymbol{b}}^{k+1} - \boldsymbol{z} + \boldsymbol{u}^{k}/\rho||_{2}^{2} \\
        & = S_{1, \lambda_{1}/\rho}(\boldsymbol{D}\tilde{\boldsymbol{b}}^{k+1} + \boldsymbol{u}^{k}/\rho).
        \label{eq:z_update}
\end{aligned}
\end{equation*}
If $\rho$ is fixed, $z^{k+1}_{r}$ will tend to be zero when $\lambda_{1} > \big|\rho\boldsymbol{D}_{r\cdot}^{T}\tilde{\boldsymbol{b}}^{k+1} + u^{k}_{r}\big|$, $r=1,\dots,J\times (M_{n}+d)$, where $\boldsymbol{D}_{r\cdot}$ is the $r$th row of $\boldsymbol{D}$.

\section{Proofs}

B-splines are essential in the estimation of coefficient function for functional model. Before presenting the proofs of the proposed estimator, it is necessary to state some properties of B-splines. As mentioned in Section 2.3, B-splines have a local support property: at most $d+1$ consecutive subintervals are nonzero. Plus, for a collection of B-spline basis functions $\{{B_{k}(t): k=1,\dots,M_{n}+d, t \in \mathcal{T}}\}$, $B_{k}(t) \geq 0$ and $\sum_{k=1}^{M_{n}+d}B_{k}(t)=1$ for all $t$. These properties imply that 
\begin{equation}
    \sup_{k,r}|\langle B_{k}, B_{r} \rangle| \leq 2(d+1)M_{n}^{-1},
\end{equation}
and thus,
\begin{equation}
    ||B_{k}||^{2}_{2} \leq \sup_{k,r}|\langle B_{k}, B_{r} \rangle| \leq 2(d+1)M_{n}^{-1}.
\end{equation} 
In addition, three inequalities will be also used. For any $x \in \mathbb{R}^{p}$
\begin{equation*}
\begin{aligned}
     ||x||_{2} & \leq ||x||_{1} \leq \sqrt{p}||x||_{2}; \\
     ||x||_{\infty} & \leq ||x||_{1} \leq p||x||_{\infty}; \\
     ||x||_{\infty} & \leq ||x||_{2} \leq \sqrt{p}||x||_{\infty}.  
\end{aligned}
\end{equation*}

\subsection{Proof of Theorem 1}

\begin{proof}
For simplicity of notation, we assume the model has no intercept, i.e., $\boldsymbol{\mu} = \boldsymbol{0}$ and rewrite the objective in Equation (2.5)
\begin{equation}
     L_{n}(\boldsymbol{b})= \frac{1}{n} ||\boldsymbol{Y} -  \boldsymbol{U}\boldsymbol{b}||_{2}^{2}  + \Delta_{n}\lambda_{1}\sum_{l=1}^{J}w^{(1)}_{l}||\boldsymbol{b}_{l}||_{1} + \lambda_{2}\sum_{l=1}^{J}w^{(2)}_{l}(\boldsymbol{b}^{T}_{l}\boldsymbol{K}_{\varphi,l}\boldsymbol{b}_{l})^{1/2},
\label{eq:obj-fadsgl-genlasso2}
\end{equation}
where $\boldsymbol{K}_{\varphi,l} = \boldsymbol{\Phi}_{l}+\varphi\boldsymbol{\Omega}_{l}$ is a $(M_{n}+d)\times(M_{n}+d)$ matrix.

We first provide Lemma 2 and 3 to facilitate the proof. Lemma 2 follows \cite{huang1} Lemma A.3 and \cite{zhou} $A_{8}$, while Lemma 3 refers to Lemma 6.2 of \cite{cardot}.

\vspace{0.3cm}
\noindent{\textbf{Lemma 2.}} \textit{If $\lim_{n \rightarrow \infty}M_{n}\log M_{n}/n=0$, there are positive constants $C_{2}$ and $C_{3}$ such that, all eigenvalues of $(M_{n}/n)\boldsymbol{U}^{T}\boldsymbol{U}$ are within the interval $[C_{2}, C_{3}]$ with probability tending to 1 as $n \rightarrow \infty$}.

\vspace{0.3cm}
\noindent{\textbf{Lemma 3.}} \textit{There are positive constants $C_{4}$ and $C_{5}$ such that,  all eigenvalues of $\boldsymbol{K}_{\varphi,l}$ are within the interval $[C_{4}\varphi M_{n}^{-1}, C_{5}M_{n}^{-1}]$ }.
\vspace{0.3cm}

\noindent{\textit{Proof}}. We define $\boldsymbol{K}_{\varphi,l} = \boldsymbol{\Phi}_{l}+\varphi\boldsymbol{\Omega}_{l}$, where $(\boldsymbol{\Phi}_{l})_{pq} = \int_{t} B_{lp}(t)B_{lq}(t)dt$ and $(\boldsymbol{\Omega}_{l})_{pq} =  \int_{t} B^{m}_{lp}(t)B^{m}_{lq} (t)dt$. The proof follows Lemma 6.2 (i) in \cite{cardot}.
\vspace{0.3cm}

Let $L_{n}(\boldsymbol{b} + \eta_{n}\boldsymbol{v}) - L_{n}(\boldsymbol{b})$, where $\eta_{n}$ is a scalar and $\boldsymbol{v} \in \mathbb{R}^{J \times (M_{n}+d)}$. At the point $\boldsymbol{b} = \boldsymbol{\alpha}$, we let $\hat{\boldsymbol{b}}= \boldsymbol{\alpha}+\eta_{n}\boldsymbol{v}$. By the minimality of $\hat{\boldsymbol{b}}$, we have
 \begin{equation}
\begin{aligned}
    & \frac{1}{n}\big(||\boldsymbol{Y} -  \boldsymbol{U}(\boldsymbol{\alpha}+\eta_{n}\boldsymbol{v})||_{2}^{2} - || \boldsymbol{Y} - \boldsymbol{U}\boldsymbol{\alpha}||_{2}^{2}\big) \\ 
   \leq  &  \Delta_{n}\lambda_{1}\sum_{l \in \mathcal{A}}w_{l}^{(1)}\big(||\boldsymbol{\alpha}_{l}||_{1} -||\boldsymbol{\alpha}_{l}+\eta_{n}\boldsymbol{v}_{l}||_{1}\big) +  \lambda_{2}\sum_{l \in \mathcal{A}}w_{l}^{(2)}\big((\boldsymbol{\alpha}_{l}^{T}\boldsymbol{K}_{\varphi,l}\boldsymbol{\alpha}_{l})^{1/2}- \\& ((\boldsymbol{\alpha}_{l}+\eta_{n}\boldsymbol{v}_{l})^{T}\boldsymbol{K}_{\varphi,l}(\boldsymbol{\alpha}_{l}+\eta_{n}\boldsymbol{v}_{l}))^{1/2}\big),
\end{aligned}
\label{eq:Dn}
\end{equation}
because of the fact that $\boldsymbol{\alpha}_{l} = \boldsymbol{0}$ if $l \in \mathcal{A}^{c}$. 

Let $\epsilon_{i} = \boldsymbol{Y}_{i} - \langle \boldsymbol{X}_{i}, \boldsymbol{\beta}\rangle$, $\boldsymbol{\epsilon} = (\epsilon_{1},\dots,\epsilon_{n})$ and $e_{i} = \langle \boldsymbol{X}_{i}, \boldsymbol{B}^{T}\boldsymbol{\alpha} \rangle - \langle \boldsymbol{X}_{i}, \boldsymbol{\beta} \rangle = \langle \boldsymbol{X}_{i}, \boldsymbol{\beta}^{\alpha}- \boldsymbol{\beta}\rangle$, $\boldsymbol{e} = (e_{1},\dots,e_{n})$, the LHS of \eqref{eq:Dn} gives
\begin{equation*}
\begin{aligned}
    \text{LHS} = & \frac{1}{n} ||\boldsymbol{\epsilon} - 
    \boldsymbol{e} - \eta_{n}\boldsymbol{U}\boldsymbol{v}||_{2}^{2} - \frac{1}{n}||\boldsymbol{\epsilon} - \boldsymbol{e}||^{2}_{2} \\
    = & \frac{\eta_{n}^{2}}{n}\boldsymbol{v}^{T}\boldsymbol{U}^{T}\boldsymbol{U}\boldsymbol{v} - \frac{2\eta_{n}}{n}(\boldsymbol{\epsilon}-\boldsymbol{e})^{T}\boldsymbol{U}\boldsymbol{v}.
\end{aligned}
\label{eq:t1o}
\end{equation*}
By Lemma 2, $(\eta_{n}^{2}/n)\boldsymbol{v}^{T}\boldsymbol{U}^{T}\boldsymbol{U}\boldsymbol{v}\geq (\eta_{n}^{2}/n)(C_{2}n/M_{n}) = \eta_{n}^{2}O_{p}(M_{n}^{-1})$.

Moreover, by Cauchy-Schwarz inequality, $(2\eta_{n}/n)(\boldsymbol{\epsilon}-\boldsymbol{e})^{T}\boldsymbol{U}\boldsymbol{v} \leq  (2\eta_{n}/n)||\boldsymbol{v}||_{2}\big((\boldsymbol{\epsilon}-\boldsymbol{e})^{T}\boldsymbol{U}\boldsymbol{U}^{T}(\boldsymbol{\epsilon}-\boldsymbol{e})\big)^{1/2}$.
Since $\boldsymbol{\epsilon}$ and $\boldsymbol{e}$ are independent,  $\mathbb{E}[(\boldsymbol{\epsilon}-\boldsymbol{e})^{T}\boldsymbol{U}\boldsymbol{U}^{T}(\boldsymbol{\epsilon}-\boldsymbol{e})^{T}] = \mathbb{E} [\boldsymbol{\epsilon}^{T}\boldsymbol{U}\boldsymbol{U}^{T}\boldsymbol{\epsilon}] + \mathbb{E}[\boldsymbol{e}^{T}\boldsymbol{U}\boldsymbol{U}^{T}\boldsymbol{e}]$.
By A.1, properties of B-splines and independence of $\epsilon_{i}$, we have
\begin{equation*}
\begin{aligned}
        \mathbb{E}[\boldsymbol{\epsilon}^{T}\boldsymbol{U}\boldsymbol{U}^{T}\boldsymbol{\epsilon}] & = \mathbb{E}\Bigg[ \sum_{l=1}^{J}\sum_{k=1}^{M_{n}+d}\Bigg[\sum_{i=1}^{n} \langle X_{li},  B_{lk} \rangle^{2}\epsilon_{i}^{2} + \sum_{i^{\prime}\neq i}\langle X_{li}, B_{lk} \rangle \langle X_{i^{\prime}}, B_{lk}\rangle \epsilon_{i}\epsilon_{li^{\prime}}\Bigg]\Bigg] \\
        & \leq \sigma^{2} \sum_{l=1}^{J}\sup_{k} \sum_{k=1}^{M_{n}+d}\sum_{i=1}^{n}|\langle X_{li}, B_{lk} \rangle|^{2} \\
        & \leq \sigma^{2}J n ||X_{li}||_{2}^{2} \sup_{k}\sum_{r=1}^{M_{n}+d}|\langle B_{lr}, B_{lk}\rangle| \\
        & = O(n).
\end{aligned}
\end{equation*}
On the other hand, by Cauchy-Schwarz inequality, A.1, and Lemma 1, $e_{i}^{2} = |\langle \boldsymbol{X}_{i}, \boldsymbol{\beta}^{\alpha}- \boldsymbol{\beta}\rangle|^{2}  \leq ||\boldsymbol{X}_{i}||_{2}^{2}\sum_{l}\int_{\mathcal{T}}\big[\sup_{t}|\beta^{\alpha}_{l}(t)-\beta_{l}(t)|\big]^{2}dt  \leq c_{1}^{2}J|\mathcal{T}|(C_{1}M_{n}^{-\delta})^{2}$,
where $|\mathcal{T}|$ represents the length of time domain. The inequality still holds for $e_{il}e_{i^{\prime}}, \forall i \neq i^{\prime}$.  Hence,
\begin{equation*}
\begin{aligned}
        \mathbb{E}[\boldsymbol{e}^{T}\boldsymbol{U}\boldsymbol{U}^{T}\boldsymbol{e}] & = \mathbb{E} \Bigg[\sum_{l=1}^{J}\sum_{k=1}^{M_{n}+d}\Bigg[\sum_{i=1}^{n} \langle X_{li}, B_{lk} \rangle^{2}e_{i}^{2} + \sum_{i^{\prime}\neq i} \langle X_{li}, B_{lk}\rangle \langle X_{li^{\prime}}, B_{lk}\rangle e_{i}e_{i^{\prime}} \Bigg]\Bigg] \\
        & \leq \sum_{k=1}^{M_{n}+d}\sum_{i=1}^{n}\  e_{i}^{2} \mathbb{E} [\langle X_{li}, B_{lk} \rangle^{2}] +  \sum_{k=1}^{M_{n}+d}\sum_{i^{\prime}\neq i}\ e_{i}e_{i^{\prime}}\mathbb{E}[\langle X_{li}, B_{lk}\rangle \langle X_{li^{\prime}}, B_{lk}\rangle] \\ 
        & = O(M^{-2\delta}_{n}nM_{n}^{-1}) + O(M^{-2\delta}_{n}n(n-1)M_{n}^{-1}) \\
        & = O(n^{2}M_{n}^{-2\delta-1}).
\end{aligned}
\end{equation*}
Therefore, by A.3, we have $\mathbb{E}[(\boldsymbol{\epsilon}-\boldsymbol{e})^{T}\boldsymbol{U}\boldsymbol{U}^{T}(\boldsymbol{\epsilon}-\boldsymbol{e})^{T}] = O(n)$. By Markov inequality, $(\boldsymbol{\epsilon}-\boldsymbol{e})^{T}\boldsymbol{U}\boldsymbol{U}^{T}(\boldsymbol{\epsilon}-\boldsymbol{e})^{T} = O_{p}(n)$ and thus,  $-(2\eta_{n}/n)(\boldsymbol{\epsilon} - \boldsymbol{e})^{T}\boldsymbol{U}\boldsymbol{v} \geq  -2\eta_{n}||\boldsymbol{v}||_{2}O_{p}(n^{-1/2})$. Therefore, the LHS gives
\begin{equation}
    \text{LHS} \geq \eta_{n}^{2}O_{p}(M_{n}^{-1}) - 2\eta_{n}O_{p}(n^{-1/2})||\boldsymbol{v}||_{2}.
    \label{eq:t1}
\end{equation}

We denote the two terms of RHS of \eqref{eq:Dn} by $T_{1}$ and $T_{2}$. We first show the convergence rate of FDoS by assuming $w^{(1)}_{l}=w^{(2)}_{l}=1$ for all $l$. With triangle inequality and $\Delta_{n} = |\mathcal{T}|/M_{n}$, and let $|\mathcal{A}|$ be the number of nonzero functions, 
\begin{equation}
\begin{aligned}
        T_{1} & \leq |\mathcal{T}|M_{n}^{-1}\lambda_{1}\eta_{n}|\mathcal{A}|(J(M_{n}+d))^{1/2}||\boldsymbol{v}||_{2} \\
         & = \eta_{n}O_{p}(\lambda_{1}M_{n}^{-1/2})||\boldsymbol{v}||_{2}.
\end{aligned}
\label{eq:t2}
\end{equation}
Suppose $\boldsymbol{x}$ is any vector and $\mathbf{A}$ is a symmetric matrix such that $g_{\mathbf{A}}
(\boldsymbol{x}) = (\boldsymbol{x}^{T}\mathbf{A}\boldsymbol{x})^{1/2}$, $g$ is a continuous function. By Taylor expansion, $g_{\boldsymbol{K}_{\varphi,l}}(\boldsymbol{\alpha}_{l}) - g_{\boldsymbol{K}_{\varphi,l}}(\boldsymbol{\alpha}_{l} + \eta_{n}\boldsymbol{v}) \leq -\eta_{n}\boldsymbol{v}^{T}\nabla g_{\boldsymbol{K}_{\varphi,l}}(\boldsymbol{\alpha}_{l})$. By Lamma 3 and $\varphi = \lambda_{2}^{2}$, $T_{2}$ gives
\begin{equation}
\begin{aligned}
        T_{2} & \leq -\lambda_{2}|\mathcal{A}|\eta_{n}(\boldsymbol{\alpha}_{l}^{T}\boldsymbol{K}_{\varphi,l}\boldsymbol{\alpha}_{l})^{-1/2}||\boldsymbol{v}||_{2}||\boldsymbol{K}_{\varphi,l}\boldsymbol{\alpha}_{l}||_{2}\mathbb{I}(\boldsymbol{\alpha}_{l} \neq \boldsymbol{0}) \\ 
        & =\eta_{n} O_{p}(\lambda_{2}^{2})||\boldsymbol{v}||_{2}.
\end{aligned}
\label{eq:t3}
\end{equation}
Combining three inequalities \eqref{eq:t1}-\eqref{eq:t3}, we have
\begin{equation*}
        \eta_{n}^{2}O_{p}(M_{n}^{-1}) - 2\eta_{n}O_{p}(n^{-1/2})||\boldsymbol{v}||_{2} \leq \eta_{n}O_{p}(\lambda_{1}M_{n}^{-1/2})||\boldsymbol{v}||_{2} + \eta_{n}  O_{p}(\lambda_{2}^{2})||\boldsymbol{v}||_{2}.
\end{equation*}
For sufficient large constant $C_{6}$ such that $||\boldsymbol{v}||_{2} = C_{6}$, we find $\eta_{n} = O_{p}(M_{n}n^{-1/2} + M_{n}^{1/2}\lambda_{1} + M_{n}\lambda_{2}^{2})$. When $\lambda_{1}=O(M_{n}^{1/2}n^{-1/2})$ and $\lambda_{2} = O(n^{-1/4})$, $\eta_{n} = O_{p}(M_{n}n^{-1/2})$. It means that for any given $\varepsilon >0$, there always exists $\eta_{n}$ such that
\begin{equation*}
    P\big\{\exists \boldsymbol{v} \in \mathbb{R}^{M_{n}+d}, ||\boldsymbol{v}||_{2}=C_{6}: L_{n}(\boldsymbol{\alpha} + \eta_{n}\boldsymbol{v}) < L_{n}(\boldsymbol{\alpha}) \big\} \geq 1 -\varepsilon.
\end{equation*}
This further means that there is a local minimizer $\hat{\boldsymbol{b}} = \boldsymbol{\alpha}+\eta_{n}\boldsymbol{v}$, such that $||\hat{\boldsymbol{b}} - \boldsymbol{\alpha}||_{2} = O_{p}(M_{n}n^{-1/2})$.

Therefore, by triangle inequality
\begin{equation*}
\begin{aligned}
    ||\hat{\boldsymbol{\beta}} - \boldsymbol{\beta}||_{\infty} & \leq ||\hat{\boldsymbol{\beta}} - \boldsymbol{\beta^{\alpha}}||_{\infty} + ||\boldsymbol{\beta}^{\alpha} - \boldsymbol{\beta}||_{\infty} \\
    & \leq \sup_{t}\sum_{k=1}^{M_{n}+d}|B_{lk}(t)|||\hat{\boldsymbol{b}} - \boldsymbol{\alpha}||_{\infty} + ||\boldsymbol{\beta}^{\alpha} - \boldsymbol{\beta}||_{\infty} \\
    & = O_{p}(M_{n}n^{-1/2}) + O(M_{n}^{-\delta})  = O_{p}(M_{n}n^{-1/2}).
\end{aligned}
\end{equation*}
The last equation holds because of A.3. 

The proof of convergence rate of FadDoS depends on the same reasoning before, except the step of $T_{1}$ and $T_{2}$. Let  $\phi_{1}=\sup_{l \in \mathcal{A}}||\check{\beta}_{l}||_{1}^{-a}$ and $\phi_{2} = \sup_{l \in \mathcal{A}}||\check{\beta}_{l}||_{2}^{-a}$. 
When $\lambda_{1}\phi_{1}=O(M_{n}^{1/2}n^{-1/2})$ and $\lambda_{2}^{2}\phi_{2} = O(n^{-1/2})$, the results follow. 
\end{proof}

\subsection{Proof of Theorem 2}

\begin{proof}
Since the penalty terms of the objective function are separable, we assume other coefficient functions fixed and only consider the $l$th coefficient function here. The overall error can be divided into estimation error and approximation error as follows.
\begin{equation*}
    (n/M_{n})^{1/2}(\hat{\beta}_{l} - \beta_{l}) = (n/M_{n})^{1/2}(\hat{\beta}_{l}-\beta^{\alpha}_{l}) + (n/M_{n})^{1/2}(\beta^{\alpha}_{l} - \beta_{l}).
    \label{eq:proof2}
\end{equation*}
Because the B-spline approximation error has been mentioned in Lemma 1, $(n/M_{n})^{1/2}(\beta^{\alpha}_{l} - \beta_{l}) = O(n^{1/2}M_{n}^{-\delta-1/2})$, we only need to focus on the first term of RHS, $(n/M_{n})^{1/2}(\hat{\beta}_{l}-\beta^{\alpha}_{l}) = (n/M_{n})^{1/2}\boldsymbol{B}_{l}^{T}(\hat{\boldsymbol{b}}_{l}-\boldsymbol{\alpha}_{l})$. Let $\boldsymbol{v} \in \mathbb{R}^{M_{n}+d}$ such that  $\hat{\boldsymbol{b}}_{l} = \boldsymbol{\alpha}_{l} + (M_{n}/n)^{1/2}\boldsymbol{v}$ and $\boldsymbol{r}_{(-l)} = \boldsymbol{Y} - \sum_{j \neq l} \boldsymbol{U}_{j}\boldsymbol{b}_{j}$, we define $Q_{n}(\boldsymbol{v})$ given \eqref{eq:obj-fadsgl-genlasso2}, assuming $w_{l}^{(1)} = w_{l}^{(2)} = 1$, 
\begin{equation*}
\begin{aligned}
    Q_{n}(\boldsymbol{v}) = & ||\boldsymbol{r}_{(-l)} - \boldsymbol{U}_{l}(\boldsymbol{\alpha}_{l} + (M_{n}/n)^{1/2}\boldsymbol{v})||_{2}^{2} + n\Delta_{n}\lambda_{1}||\boldsymbol{\alpha}_{l}+(M_{n}/n)^{1/2}\boldsymbol{v}||_{1} +  \\ 
    & n\lambda_{2}((\boldsymbol{\alpha}_{l}+(M_{n}/n)^{1/2}\boldsymbol{v})^{T}\boldsymbol{K}_{\varphi,l}(\boldsymbol{\alpha}_{l}+(M_{n}/n)^{1/2}\boldsymbol{v}))^{1/2}.
\end{aligned}
\end{equation*}
Suppose the minimizer of $Q_{n}(\boldsymbol{v})$ is noted as $\hat{\boldsymbol{v}}_{n}$, then $\hat{\boldsymbol{v}}_{n} = (n/M_{n})^{1/2}(\hat{\boldsymbol{b}}_{l} - \boldsymbol{\alpha}_{l})$. We need to show the limiting distribution of $\hat{\boldsymbol{v}}_{n}$ by proving the finite distribution convergence of $V_{1(n)}^{(l)}$ to $V_{1}^{(l)}$. Note that $V^{(l)}_{1(n)}(\boldsymbol{v})=Q_{n}(\boldsymbol{v})-Q_{n}(\boldsymbol{0})$,
\begin{equation*}
 \begin{aligned}
     V_{1(n)}^{(l)}(\boldsymbol{v}) =
   &  [\boldsymbol{v}^{T}(\frac{M_{n}}{n}\boldsymbol{U}_{l}^{T} \boldsymbol{U}_{l})\boldsymbol{v} - 2(\frac{M_{n}}{n})^{1/2}(\boldsymbol{r}_{(-l)} -\boldsymbol{U}_{l}\boldsymbol{\alpha}_{l})^{T}\boldsymbol{U}_{l}\boldsymbol{v}]  + \\
   & n\Delta_{n}\lambda_{1}\big(||\boldsymbol{\alpha}_{l}+(M_{n}/n)^{1/2}\boldsymbol{v}||_{1} -  ||\boldsymbol{\alpha}_{l}||_{1}\big) + \\ 
   & n\lambda_{2}\big(((\boldsymbol{\alpha}_{l}+(M_{n}/n)^{1/2}\boldsymbol{v})^{T}\boldsymbol{K}_{\varphi,l}(\boldsymbol{\alpha}_{l}+(M_{n}/n)^{1/2}\boldsymbol{v}))^{1/2} - (\boldsymbol{\alpha}^{T}_{l}\boldsymbol{K}_{\varphi,l}\boldsymbol{\alpha}_{l})^{1/2}\big).
\end{aligned} 
\end{equation*} 
Because $(M_{n}/n)\boldsymbol{U}_{l}^{T} \boldsymbol{U}_{l} \rightarrow \boldsymbol{C}_{l}$ and $\boldsymbol{W}_{l} \sim N(\boldsymbol{0},\sigma^{2}\boldsymbol{C}_{l})$, the first term denoted by $T_{1}$ of RHS of $V_{1(n)}^{(l)}(\boldsymbol{v})$ is 
\begin{equation*}
   T_{1} = \boldsymbol{v}^{T}\boldsymbol{C}_{l}\boldsymbol{v} - 2\boldsymbol{W}_{l}^{T}\boldsymbol{v},
\end{equation*}
which is due to 
\begin{equation*}
(M_{n}/n)^{1/2}(\boldsymbol{r}_{(-l)} -  \boldsymbol{U}_{l}\boldsymbol{\alpha}_{l})^{T}\boldsymbol{U}_{l} =  (M_{n}/n)^{1/2}\boldsymbol{\epsilon}^{T}\boldsymbol{U}_{l} + (M_{n}/n)^{1/2}\boldsymbol{e}^{T}\boldsymbol{U}_{l}.
\end{equation*}
We can see that $(M_{n}/n)^{1/2}\boldsymbol{\epsilon}^{T}\boldsymbol{U}_{l} = \boldsymbol{W}_{l}$ and $(M_{n}/n)^{1/2}\boldsymbol{e}^{T}\boldsymbol{U}_{l} = (M_{n}/n)^{1/2}O_{p}(M_{n}^{-\delta-1/2}n) =  o_{p}(1)$. Furthermore, the second term denoted by $T_{2}$ of RHS is 
\begin{equation*}
    T_{2} = \lambda_{1}(n/M_{n})^{1/2}\sum_{k=1}^{M_{n}+d}\bigg\{
    |v_{k}|\mathbb{I}(\alpha_{lk} = 0) + v_{k}\text{sgn}\big(\alpha_{lk}\big) \mathbb{I}(\alpha_{lk} \neq 0) \bigg\}.
\end{equation*}
By Lemma 3, the eigenvalues of $\boldsymbol{K}_{\varphi,l}$ is of the order $O(\varphi M_{n}^{-1})$, the third term denoted by $T_{3}$ of RHS gives 
\begin{equation*}
\begin{aligned}
     T_{3} & = \lambda_{2}(M_{n}n)^{1/2}\bigg\{(\boldsymbol{\alpha}_{l}^{T}\boldsymbol{K}_{\varphi,l}\boldsymbol{\alpha}_{l})^{-1/2}\boldsymbol{v}^{T} \boldsymbol{K}_{\varphi,l}\boldsymbol{\alpha}_{l} \mathbb{I}(\boldsymbol{\alpha}_{l} \neq \boldsymbol{0}) + (\boldsymbol{v}^{T}\boldsymbol{K}_{\varphi,l}\boldsymbol{v})^{1/2}\mathbb{I}(\boldsymbol{\alpha}_{l} = \boldsymbol{0}) \bigg\} + o(M_{n})\lambda_{2} \\
     &= \lambda_{2}^{2}n^{1/2}\bigg\{||\boldsymbol{v}||_{2}\mathbb{I}(\boldsymbol{\alpha}_{l} = \boldsymbol{0}) + (\boldsymbol{v}^{T}\boldsymbol{\alpha}_{l}/||\boldsymbol{\alpha}_{l}||_{2}) \mathbb{I}(\boldsymbol{\alpha}_{l} \neq \boldsymbol{0}) \bigg\} + o(M_{n})\lambda_{2}.   
\end{aligned}
\end{equation*}

So, we have $V^{(l)}_{1(n)}(\boldsymbol{v}) \overset{p}{\rightarrow} V^{(l)}_{1}(\boldsymbol{v})$ for a fixed $\boldsymbol{v}$. $V^{(l)}_{1(n)}$ is a convex function and it follows the results of \cite{geyer} that $(n/M_{n})^{1/2}(\hat{\boldsymbol{b}}_{l} - \boldsymbol{\alpha}_{l}) = \hat{\boldsymbol{v} }_{n} = \argmin_{\boldsymbol{v}} V^{(l)}_{1(n)} \overset{p}{\rightarrow} \argmin_{\boldsymbol{v}} V^{(l)}_{1}$ for $l=1,\dots,J$. Multipling $\boldsymbol{B}_{l}(t)$ on both sides obtains
\begin{equation*}
\begin{aligned}
        (n/M_{n})^{1/2}(\hat{\beta}_{l}(t) - \beta_{l}(t)) & = (n/M_{n})^{1/2}(\hat{\beta}_{l}(t)-\beta^{\alpha}_{l}(t)) + O(n^{1/2}M_{n}^{-\delta-1/2}) \\
        & \overset{d}{\rightarrow} \boldsymbol{B}_{l}^{T}(t)\argmin_{\boldsymbol{v}}V^{(l)}_{1}(\boldsymbol{v}).
\end{aligned} 
\end{equation*} 
\end{proof}

\subsection{Proof of Proposition 1}

\begin{proof}[Proof of 1.]
Since finding $\hat{\mathcal{A}}_{n} = \mathcal{A}$ is the intersection of correctly estimating nonzero values for nonzero coefficient function and correctly identifying zero coefficient function,  
$P(\hat{\mathcal{A}}_{n}=\mathcal{A}) \leq P(\hat{\boldsymbol{b}}_{l} = \boldsymbol{0} \ \forall l \notin \mathcal{A})$. 
For $l$th coefficient functions, let $\boldsymbol{v}^{*}_{l} = \argmin_{\boldsymbol{v}} V_{1}^{(l)}(\boldsymbol{v}_{l})$, Theorem 2 shows that $(n/M_{n})^{1/2}(\hat{\boldsymbol{b}}_{l} - \boldsymbol{\alpha}_{l}) \overset{d}{\rightarrow} \boldsymbol{v}^{*}_{l}$. Therefore, we need to show that $c = P(\boldsymbol{v}^{*}_{l} = \boldsymbol{0} \ \forall l \notin \mathcal{A}) < 1$. There
are two cases:

If $\gamma_{1}=\gamma_{2}=0$, $\boldsymbol{v}^{*}_{l}=\boldsymbol{C}^{-1}_{l}\boldsymbol{W}_{l} \sim N(\boldsymbol{0}, \sigma^{2}\boldsymbol{C}^{-1}_{l})$ and therefore $c=0$.

If $\gamma_{1} \neq 0$ or $\gamma_{2} \neq 0$,
\begin{equation*}
V_{1}^{(l)}(\boldsymbol{v}_{l}) =
\begin{cases}
  \boldsymbol{v}_{l}^{T}\boldsymbol{C}_{l}\boldsymbol{v}_{l} - 2\boldsymbol{W}_{l}^{T}\boldsymbol{v}_{l} + \gamma_{1}\Gamma_{1}^{(l)}(\boldsymbol{v}_{l}) + \gamma_{2} (\boldsymbol{v}_{l}
  ^{T}\boldsymbol{\alpha}_{l}/||\boldsymbol{\alpha}_{l}||_{2}) & \text{if} \ l \in \mathcal{A} \\

  \boldsymbol{v}_{l}^{T}\boldsymbol{C}_{l}\boldsymbol{v}_{l} - 2\boldsymbol{W}_{l}^{T}\boldsymbol{v}_{l} +  \gamma_{1}||\boldsymbol{v}_{l}||_{1} +  \gamma_{2}||\boldsymbol{v}_{l}||_{2} & \text{if} \ l \notin \mathcal{A}, 
\end{cases}
\end{equation*}
It can be seen that $V_{1}^{(l)}(\boldsymbol{v}_{l})$ is not differentiable at $v_{lk}=0$ for any $k$. By KKT conditions, $\boldsymbol{v}^{*}_{l}$ should satisfies
\begin{equation}
\begin{cases}
    2\boldsymbol{C}_{l}\boldsymbol{v}^{*}_{l}-2\boldsymbol{W}_{l} + \gamma_{1}\boldsymbol{p}_{l} + \gamma_{2}\boldsymbol{\alpha}_{l}/||\boldsymbol{\alpha}_{l}||_{2} = 0  & \text{if} \ l \in \mathcal{A} \\
    2\boldsymbol{C}_{l}\boldsymbol{v}^{*}_{l}-2\boldsymbol{W}_{l} + \gamma_{1}\boldsymbol{q}_{l} + \gamma_{2}\boldsymbol{z}_{l} = 0  & \text{if} \ l \notin \mathcal{A},
\end{cases}
\label{cond:kkt}
\end{equation}
where $p_{lk}=\partial_{v_{lk}^{*}} \big\{
    |v_{lk}^{*}|\mathbb{I}(\alpha_{lk} = 0) + v_{lk}^{*}\text{sgn}\big(\alpha_{lk}\big) \mathbb{I}(\alpha_{lk} \neq 0) \big\}$,
\begin{equation*}
q_{lk} =  
\begin{cases}
    \text{sgn}(v_{lk}^{*}) & \text{if} \ v_{lk}^{*} \neq 0 \\
    \in \{q_{lk}: |q_{lk}| \leq 1 \} & \text{if} \ v_{lk}^{*} = 0, \\
\end{cases}
\hspace{0.3cm}
\boldsymbol{z}_{l} =
\begin{cases}
    \boldsymbol{v}^{*}_{l}/||\boldsymbol{v}^{*}_{l}||_{2} & \text{if} \ \boldsymbol{v}^{*}_{l} \neq \boldsymbol{0} \\
    \in \{\boldsymbol{z}_{l}: ||\boldsymbol{z}_{l}||_{2} \leq 1 \} & \text{if} \ \boldsymbol{v}^{*}_{l} = \boldsymbol{0}.
\end{cases}
\end{equation*}

To examine the variable selection consistency, we have to introduce new denotations by combining all coefficient functions. We let $\boldsymbol{C} = (M_{n}/n) \boldsymbol{U}^{T}\boldsymbol{U}$, $\boldsymbol{U} = (\boldsymbol{U}_{1},\dots,\boldsymbol{U}_{J})^{T}$, $\boldsymbol{W} = (\boldsymbol{W}_{1},\dots,\boldsymbol{W}_{J})^{T}$, $\boldsymbol{p} = (\boldsymbol{p}_{1}, \dots, \boldsymbol{p}_{J})^{T}$, $\boldsymbol{q} = (\boldsymbol{q}_{1}, \dots, \boldsymbol{q}_{J})^{T}$, $\boldsymbol{\alpha} = (\boldsymbol{\alpha}_{1}, \dots, \boldsymbol{\alpha}_{J})^{T}$, and $\boldsymbol{v}^{*} = (\boldsymbol{v}_{1}^{*}, \dots, \boldsymbol{v}_{J}^{*})^{T}$. Without loss of generality, rewrite matrix $\boldsymbol{C}$ in a block-wise form involving either $l \in \mathcal{A}$ or $\mathcal{A}^{c}$ such that  
\begin{equation*}
\boldsymbol{C} =
\begin{bmatrix}
    \boldsymbol{C}_{\mathcal{A}\mathcal{A}}  &  \boldsymbol{C}_{\mathcal{A}\mathcal{A}^{c}} \\
     \boldsymbol{C}_{\mathcal{A}^{c}\mathcal{A}}  &  \boldsymbol{C}_{\mathcal{A}^{c}\mathcal{A}^{c}} \\
\end{bmatrix},
\end{equation*}
and rewrite vectors $\boldsymbol{W}$, $\boldsymbol{p}$
, $\boldsymbol{q}$, $\boldsymbol{\alpha}$, and $\boldsymbol{v}^{*}_{l}$ likewise.
If $\boldsymbol{v}^{*}_{l}=\boldsymbol{0}$ for any $l \notin \mathcal{A}$, the optimality conditions \eqref{cond:kkt} become
\begin{equation}
\begin{cases}
2\boldsymbol{C}_{\mathcal{A}\mathcal{A}}\boldsymbol{v}_{\mathcal{A}}^{*}-2\boldsymbol{W}_{\mathcal{A}} + \gamma_{1}\boldsymbol{p}_{\mathcal{A}}  + \gamma_{2}\boldsymbol{\alpha}_{\mathcal{A}}/||\boldsymbol{\alpha}_{\mathcal{A}}||_{2} = 0 \\
||2\boldsymbol{C}_{\mathcal{A}^{c}\mathcal{A}}\boldsymbol{v}_{\mathcal{A}}^{*} -2\boldsymbol{W}_{\mathcal{A}^{c}} + \gamma_{1}\boldsymbol{q}_{\mathcal{A}^{c}}||_{2} \leq \gamma_{2}.
\end{cases}
\label{cond:kkt1}
\end{equation}
Combining these optimality conditions \eqref{cond:kkt1} above, 
\begin{equation*}
\Big|\Big|\boldsymbol{C}_{\mathcal{A}^{c}\mathcal{A}}\boldsymbol{C}_{\mathcal{A}\mathcal{A}}^{-1} (2\boldsymbol{W}_{\mathcal{A}} - \gamma_{1}\boldsymbol{p}_{\mathcal{A}}  - \gamma_{2}\boldsymbol{\alpha}_{\mathcal{A}}/||\boldsymbol{\alpha}_{\mathcal{A}}||_{2}) - 2\boldsymbol{W}_{\mathcal{A}^{c}}  + \gamma_{1}\boldsymbol{q}_{\mathcal{A}^{c}} \Big|\Big|_{2}\leq \gamma_{2}.
\end{equation*}
Thus, we obtain, 
\begin{equation*}
    c \leq P\bigg\{\Big|\Big|\boldsymbol{C}_{\mathcal{A}^{c}\mathcal{A}}\boldsymbol{C}_{\mathcal{A}\mathcal{A}}^{-1} (2\boldsymbol{W}_{\mathcal{A}} - \gamma_{1}\boldsymbol{p}_{\mathcal{A}}  - \gamma_{2}\boldsymbol{\alpha}_{\mathcal{A}}/||\boldsymbol{\alpha}_{\mathcal{A}}||_{2}) - 2\boldsymbol{W}_{\mathcal{A}^{c}} + \gamma_{1}\boldsymbol{q}_{\mathcal{A}^{c}} \Big|\Big|_{2}\leq \gamma_{2} \bigg\} < 1.
\end{equation*}
\end{proof}

\begin{proof}[Proof of 2] Given the same reasoning of 1, $P(\hat{\beta}_{l}(t)=0) \leq P(\hat{b}_{lk} = \boldsymbol{0} \ \forall k \notin \mathcal{B})$. 
We need to show that $c = P(v^{*}_{lk} = \boldsymbol{0} \ \forall k \notin \mathcal{B}) < 1$. There are also two cases:

If $\gamma_{1}=\gamma_{2}=0$, $\boldsymbol{v}^{*}_{l}=\boldsymbol{C}^{-1}_{l}\boldsymbol{W}_{l} \sim N(\boldsymbol{0}, \sigma^{2}\boldsymbol{C}^{-1}_{l})$ and therefore $c=0$.

If $\gamma_{1}=\gamma_{2}=0$, for $l \in \mathcal{A}$, 
\begin{equation*}
\begin{aligned}
    V_{1}^{(l)}(\boldsymbol{v}_{l}) =  &
  \boldsymbol{v}_{l}^{T}\boldsymbol{C}_{l}\boldsymbol{v}_{l} - 2\boldsymbol{W}_{l}\boldsymbol{v}_{l} + \gamma_{1}\Gamma_{1}^{(l)}(\boldsymbol{v}_{l})
    + \gamma_{2} \boldsymbol{v}_{l}^{T} \boldsymbol{\alpha}_{\mathcal{A}}/||\boldsymbol{\alpha}_{\mathcal{A}}||_{2} \\
     = &
    \begin{cases}
  \boldsymbol{v}_{l}^{T}\boldsymbol{C}_{l}\boldsymbol{v}_{l} - 2\boldsymbol{W}_{l}\boldsymbol{v}_{l} + \gamma_{1}v_{lk}\text{sgn}\big(\alpha_{lk}\big) +
    \gamma_{2} \boldsymbol{v}_{l}^{T} \boldsymbol{\alpha}_{\mathcal{A}}/||\boldsymbol{\alpha}_{\mathcal{A}}||_{2} & \text{if} \ k \in \mathcal{B} \\

  \boldsymbol{v}_{l}^{T}\boldsymbol{C}_{l}\boldsymbol{v}_{l} - 2\boldsymbol{W}_{l}\boldsymbol{v}_{l} +  \gamma_{1}|v_{lk}| +  \gamma_{2} \boldsymbol{v}_{l}^{T} \boldsymbol{\alpha}_{\mathcal{A}}/||\boldsymbol{\alpha}_{\mathcal{A}}||_{2} & \text{if} \ k \notin \mathcal{B}. 
    \end{cases}
\end{aligned}
\end{equation*}
We can follow the KKT conditions with respect to individual $v_{lk}$, and then depend on the similar reasoning of the \textit{Proof of 1.} to obtain the results.
\end{proof}

\subsection{Proof of Theorem 3}

\begin{proof}
The proof of Theorem 3 requires Lemma 4, which shows the rate of convergence of the initial estimator which is derived from the penalized B-splines estimator in \cite{cardot}. Let $\lambda$ be the tuning parameter for the functional generalization of ridge regularization.

\vspace{0.3cm}
\noindent{\textbf{Lemma 4.}} \textit{Under} (A.1)-(A.3), \textit{if $\lambda=O(M_{n}n^{-1/2})$, $||\hat{\check{\beta}}_{l} - \beta_{l}||_{\infty} = O_{p}(M_{n}n^{-1/2})$}.
\noindent{\textit{Proof}}. The proof depends on the identical procedure as Theorem 1. 
\vspace{0.3cm}

Now we are ready to give the proof of Theorem 3. Based on \eqref{eq:obj-fadsgl-genlasso2}, we define $Q_{n}(\boldsymbol{v})$, $\boldsymbol{v} \in \mathbb{R}^{M_{n} +d}$ by adding adaptive weights as following
\begin{equation*}
\begin{aligned}
    Q_{n}(\boldsymbol{v}) = & ||\boldsymbol{r}_{(-l)} - \boldsymbol{U}_{l}(\boldsymbol{\alpha}_{l} + (M_{n}/n)^{1/2}\boldsymbol{v})||_{2}^{2} + n\Delta_{n}\lambda_{1}\hat{w}^{(1)}_{l}||\boldsymbol{\alpha}_{l}+(M_{n}/n)^{1/2}\boldsymbol{v}||_{1} + \\
    & n\lambda_{2}\hat{w}^{(2)}_{l}((\boldsymbol{\alpha}_{l}+(M_{n}/n)^{1/2}\boldsymbol{v})^{T}\boldsymbol{K}_{\varphi,l}(\boldsymbol{\alpha}_{l}+(M_{n}/n)^{1/2}\boldsymbol{v}))^{1/2}.
\end{aligned}
\end{equation*}
Suppose the minimizer of $Q_{n}(\boldsymbol{v})$ is noted as $\hat{\boldsymbol{v}}_{n}$, then $\hat{\boldsymbol{v}}_{n} = (n/M_{n})^{1/2}(\hat{\boldsymbol{b}}_{l} - \boldsymbol{\alpha}_{l})$. We need to show the limiting distribution of $\hat{\boldsymbol{v}}_{n}$. Similarly, $V^{(l)}_{2(n)}(\boldsymbol{v})=Q_{n}(\boldsymbol{v})-Q_{n}(\boldsymbol{0}) = \boldsymbol{v}^{T}\boldsymbol{C}_{l}\boldsymbol{v} - 2\boldsymbol{W}_{l}^{T}\boldsymbol{v} + \Gamma_{1,2}(\boldsymbol{v})$, where
\begin{equation*}
\Gamma_{1,2}(\boldsymbol{v}) =
\begin{cases}
   \lambda_{1}(n/M_{n})^{1/2}\hat{w}^{(1)}_{l}\sum_{k=1}^{M_{n}+d}\bigg\{
    |v_{k}|\mathbb{I}(\alpha_{k} = 0) +   v_{k}\text{sgn}\big(\alpha_{lk}\big) \mathbb{I}(\alpha_{lk} \neq 0) \bigg\} + \\
    \lambda_{2}(M_{n}n)^{1/2}\hat{w}^{(2)}_{l}  (\boldsymbol{\alpha}_{l}^{T}\boldsymbol{K}_{\varphi,l}\boldsymbol{\alpha}_{l})^{-1/2}\boldsymbol{v}^{T}\boldsymbol{K}_{\varphi,l}\boldsymbol{\alpha}_{l} & \text{if} \ l \in \mathcal{A} \\

    \lambda_{1}(n/M_{n})^{1/2}\hat{w}^{(1)}_{l}||\boldsymbol{v}||_{1} +  \lambda_{2}(M_{n}n)^{1/2}\hat{w}^{(2)}_{l}(\boldsymbol{v}^{T}\boldsymbol{K}_{\varphi,l}\boldsymbol{v})^{1/2} & \text{if} \ l \notin \mathcal{A}, \\
\end{cases}
\end{equation*}
and $(M_{n}/n)\boldsymbol{U}_{l}^{T} \boldsymbol{U}_{l} \rightarrow \boldsymbol{C}_{l}$,
$\boldsymbol{W}_{l}=N(\boldsymbol{0},\sigma^{2}\boldsymbol{C}_{l})$ as before. 

If $l \notin \mathcal{A}$, $\hat{w}^{(1)}_{l} = ||\hat{\check{\beta}}||_{1}^{-a} = O_{p}(M_{n}^{-a}n^{a/2})$ and $\hat{w}^{(2)}_{l} = ||\hat{\check{\beta}}||_{2}^{-a} = O_{p}(M_{n}^{-a}n^{a/2})$,
\begin{equation*}
\lambda_{1}(\frac{n}{M_{n}})^{1/2}\hat{w}^{(1)}_{l}||\boldsymbol{v}||_{1} \propto \lambda_{1}(\frac{n}{M_{n}})^{1/2}\frac{n^{a/2}}{M_{n}^{a}}\hat{w}^{(1)}_{l}\frac{M_{n}^{a}}{n^{a/2}}||\boldsymbol{v}||_{1} = \frac{\lambda_{1}n^{(a+1)/2}}{M_{n}^{a + 1/2}}\hat{w}^{(1)}_{l}\frac{M_{n}^{a}}{n^{a/2}}||\boldsymbol{v}||_{1}  \overset{p}{\rightarrow} \infty,
\end{equation*}
since $\lambda_{1}n^{(a+1)/2}/M_{n}^{a+1/2} \rightarrow \infty$ and $(M_{n}^{a}/n^{a/2})\hat{w}^{(1)}_{l} = O_{p}(1)$; on the other hand, given Lemma 3 and $\varphi = \lambda_{2}^{2}$,
\begin{equation*}
    \lambda_{2}(M_{n}n)^{1/2}\hat{w}^{(2)}_{l}(\boldsymbol{v}^{T}\boldsymbol{K}_{\varphi,l}\boldsymbol{v})^{1/2} = \lambda_{2}^{2}n^{1/2}\frac{n^{a/2}}{M_{n}^{a}}\hat{w}^{(2)}_{l}\frac{M_{n}^{a}}{n^{a/2}} = \frac{\lambda_{2}^{2}n^{(a+1)/2}}{M_{n}^{a}}\hat{w}^{(2)}_{l}\frac{M_{n}^{a}}{n^{a/2}} \overset{p}{\rightarrow} \infty,
\end{equation*}
since $\lambda_{2}^{2}n^{(a+1)/2}/M_{n}^{a} \rightarrow \infty$ and $(M_{n}^{a}/n^{a/2})\hat{w}^{(2)}_{l} = O_{p}(1)$. Hence, for $l \notin \mathcal{A}$
\begin{equation}
    V_{2(n)}^{(l)}(\boldsymbol{v}) = \infty.
\label{eq:4.1}
\end{equation}

If $l \in \mathcal{A}$ and $k \notin \mathcal{B}$, we have 
\begin{equation*}
    V_{2(n)}^{(l)}(\boldsymbol{v}) = \boldsymbol{v}^{T}\boldsymbol{C}_{l}\boldsymbol{v} - 2\boldsymbol{W}_{l}^{T}\boldsymbol{v} + \lambda_{1}(\frac{n}{M_{n}})^{1/2}\hat{w}^{(1)}_{l}|v_{k}| +  \\ \lambda_{2}(M_{n}n)^{1/2}\hat{w}^{(2)}_{l}  (\boldsymbol{\alpha}_{l}^{T}\boldsymbol{K}_{\varphi,l}\boldsymbol{\alpha}_{l})^{-1/2}\boldsymbol{v}^{T}\boldsymbol{K}_{\varphi,l}\boldsymbol{\alpha}_{l}.
\end{equation*}
Under the same conditions before, we know that $\lambda_{1}(\frac{n}{M_{n}})^{1/2}\hat{w}^{(1)}_{l}|v_{k}| \overset{p}{\rightarrow} \infty$ and \\ $\lambda_{2}(M_{n}n)^{1/2}\hat{w}^{(2)}_{l} (\boldsymbol{\alpha}_{l}^{T}\boldsymbol{K}_{\varphi,l}\boldsymbol{\alpha}_{l})^{-1/2}\boldsymbol{v}^{T} \boldsymbol{K}_{\varphi,l}\boldsymbol{\alpha}_{l} \overset{p}{\rightarrow} \infty$. We re-expressed $\boldsymbol{v}$ in a blockwise form mentioned before such that $\boldsymbol{v} = (\boldsymbol{v}_{\mathcal{B}}, \boldsymbol{v}_{\mathcal{B}^{c}})^{T}$.
Hence,
\vspace{-0.4cm}
\begin{equation}
    V_{2(n)}^{(l)}(\boldsymbol{v}_{\mathcal{B}^{c}}) = \infty.
\label{eq:4.2}
\end{equation}

If $l \in \mathcal{A}$ and $k \in \mathcal{B}$, due to $t \in I_{0}(\beta_{l})$, $\hat{\check{\beta}}_{l}(t)  \overset{p}{\rightarrow} \beta_{l}(t) \neq 0$, the last two terms converge to zero in probability under the conditions $\lambda_{1}(n/M_{n})^{1/2} \rightarrow 0$ and $\lambda_{2}^{2}n^{1/4} \rightarrow 0$. Thus,
\begin{equation}
    V_{2(n)}^{(l)}(\boldsymbol{v}_{\mathcal{B}}) = \boldsymbol{v}_{\mathcal{B}}^{T}(\boldsymbol{C}_{l})_{\mathcal{B}\mathcal{B}}\boldsymbol{v}_{\mathcal{B}} - 2(\boldsymbol{W}_{l})_{\mathcal{B}}^{T}\boldsymbol{v}_{\mathcal{B}}.
\label{eq:4.3}
\end{equation}

Thus, we see that $V^{(l)}_{2(n)}(\boldsymbol{v}) \overset{p}{\rightarrow} V^{(l)}_{2}(\boldsymbol{v})$ for a fixed $\boldsymbol{v}$ and $V^{(l)}_{2(n)}$ is a convex function. Given \eqref{eq:4.1}, \eqref{eq:4.2}, and \eqref{eq:4.3}, it follows the results of \cite{geyer} that $\hat{\boldsymbol{v}}_{n}^{(l)} = \argmin_{\boldsymbol{v}} V^{(l)}_{2(n)} \overset{p}{\rightarrow} \argmin_{\boldsymbol{v}} V^{(l)}_{2}$,
where 
\begin{equation*}
V^{(l)}_{2}(\boldsymbol{v}) =
\begin{cases}
   \boldsymbol{v}_{\mathcal{B}}(\boldsymbol{C}_{l})_{\mathcal{B}\mathcal{B}}\boldsymbol{v}_{\mathcal{B}} - 2(\boldsymbol{W}_{l})^{T}_{\mathcal{B}}\boldsymbol{v}_{\mathcal{B}} & \text{if} \ l \in \mathcal{A} \ \text{and} \ k \in \mathcal{B} \\
    \infty & \text{if} \ l \in \mathcal{A} \ \text{and} \ k \notin \mathcal{B} \\
  \infty & \text{if} \ l \notin \mathcal{A}, \\
\end{cases}
\end{equation*}
and $(\boldsymbol{C}_{l})_{\mathcal{B}\mathcal{B}}$,  $(\boldsymbol{W}_{l})_{\mathcal{B}}$ are similarly defined. 
Therefore, we have $\hat{\boldsymbol{v}}_{n} \overset{d}{\rightarrow} (\boldsymbol{C}_{l})_{\mathcal{B}\mathcal{B}}^{-1}(\boldsymbol{W}_{l})_{\mathcal{B}}$ if $l \in \mathcal{A}$ and $k \in \mathcal{B}$; $\hat{\boldsymbol{v}}_{n}\overset{d}{\rightarrow} \boldsymbol{0}$ if $l \in \mathcal{A}$ and $k \notin \mathcal{B}$; $\hat{\boldsymbol{v}}_{n} \overset{d}{\rightarrow} \boldsymbol{0}$ if $l \notin \mathcal{A}$.
Since $(\boldsymbol{W}_{l})_{\mathcal{B}} \sim N(\boldsymbol{0}, \sigma^{2}(\boldsymbol{C}_{l})_{\mathcal{B}\mathcal{B}})$, it can be seen that for $t \in I_{1}(\beta_{l})$,
\begin{equation*}
\begin{aligned}
        (n/M_{n})^{1/2}(\hat{\beta}_{l}(t) - \beta_{l}(t)) & = (n/M_{n})^{1/2}(\hat{\beta}_{l}(t)-\beta^{\alpha}_{l}(t)) + (n/M_{n})^{1/2}(\beta^{\alpha}_{l}(t)-\beta_{l}(t)) \\
        & = (n/M_{n})^{1/2}\boldsymbol{B}_{l}^{T}(\hat{\boldsymbol{b}}_{l\mathcal{B}} - \boldsymbol{\alpha}_{l\mathcal{B}}) + O(n^{1/2}M_{n}^{-\delta-1/2}) \\
        & \overset{d}{\rightarrow} \boldsymbol{B}_{l}^{T}(\boldsymbol{C}_{l})_{\mathcal{B}\mathcal{B}}^{-1}(\boldsymbol{W}_{l})_{\mathcal{B}} \\
        & \overset{d}{\rightarrow} N(0, \Sigma_{lt}), 
\end{aligned} 
\end{equation*} 
where $\Sigma_{lt} = \sigma^{2}\boldsymbol{B}_{l}^{T}(t)(\boldsymbol{C}_{l})_{\mathcal{B}\mathcal{B}}^{-1}\boldsymbol{B}_{l}(t)$.

We now want to prove the consistency of global and local selection. We start with the global variable selection. The asymptotic normality indicates that $P(l \in \hat{\mathcal{A}}) \rightarrow 1$. Then it suffices to show that for any $l^{\prime} \notin \mathcal{A}$, $P(l^{\prime} \in \hat{\mathcal{A}}) \rightarrow 0$. We rewrite the objective function for $\boldsymbol{b}_{l^{\prime}}$, resulting in 
$L_{n}(\boldsymbol{b}_{l^{\prime}})=  n^{-1}||\boldsymbol{r}_{(-l^{\prime})} -  \boldsymbol{U}_{l^{\prime}}\boldsymbol{b}_{l^{\prime}}||_{2}^{2}  + \Delta_{n}\lambda_{1}\hat{w}^{(1)}_{l^{\prime}}||\boldsymbol{b}_{l^{\prime}}||_{1} + \lambda_{2}\hat{w}^{(2)}_{l^{\prime}} (\boldsymbol{b}_{l^{\prime}}^{T}\boldsymbol{K}_{\varphi,l^{\prime}}\boldsymbol{b}_{l^{\prime}})^{1/2}$. By Lemma 3 that eigenvalues of $\boldsymbol{K}_{\varphi,l}$ is of the order $O(\varphi M_{n}^{-1})$, we can derive the KKT conditions for global and local selection respectively.

If $l^{\prime} \in \hat{\mathcal{A}}$, under KKT conditions, we have
\begin{equation}
       2\boldsymbol{U}_{l^{\prime}}^{T}(\boldsymbol{r}_{(-l^{\prime})} - \boldsymbol{U}_{l^{\prime}}\boldsymbol{b}_{l^{\prime}}) = n\Delta_{n}\lambda_{1}\hat{w}^{(1)}_{l^{\prime}}\boldsymbol{q}_{l^{\prime}} + nM_{n}^{-1/2}\lambda_{2}^{2}\hat{w}^{(2)}_{l^{\prime}}\boldsymbol{b}_{l^{\prime}}/||\boldsymbol{b}_{l^{\prime}}||_{2}
\label{eq:kkt_faddos1}
\end{equation}
where $q_{l^{\prime}k}$ is defined similary as above with respect to $b_{l^{\prime}k}$ here. Multiplying $(M_{n}/n)^{1/2}$ on both sides, the LHS gives
\begin{equation*}
\begin{aligned}
        2(\frac{M_{n}}{n})^{1/2}\boldsymbol{U}_{l^{\prime}}^{T}(\boldsymbol{r}_{(-l^{\prime})} - \boldsymbol{U}_{l^{\prime}}\boldsymbol{b}_{l^{\prime}}) 
        = 2\boldsymbol{W}_{l^{\prime}} + 2(\frac{M_{n}}{n})\boldsymbol{U}_{l^{\prime}}^{T}\boldsymbol{U}_{l^{\prime}}(\frac{n}{M_{n}})^{1/2}(\boldsymbol{\alpha}_{l^{\prime}}-\boldsymbol{b}_{l^{\prime}}).
\end{aligned}
\end{equation*}
Because $\boldsymbol{W}_{l^{\prime}} \sim N(\boldsymbol{0}, \sigma^{2}\boldsymbol{C}_{l^{\prime}})$ and the second term asymptotically converges to normal distribution, $2(M_{n}/n)^{1/2}\boldsymbol{U}_{l^{\prime}}^{T}(\boldsymbol{r}_{(-l^{\prime})} - \boldsymbol{U}_{l^{\prime}}\boldsymbol{b}_{l^{\prime}}) \overset{d}{\rightarrow} \text{some normal distribution}$ by Slutsky's theorem. Given $(n/M_{n})^{1/2}\lambda_{1}\hat{w}^{(1)}_{l} \overset{p}{\rightarrow} \infty$ and $n^{1/2}\lambda_{2}^{2}\hat{w}^{(2)}_{l} \overset{p}{\rightarrow} \infty$ under the conditions, we know the probability that \eqref{eq:kkt_faddos1} holds tends to be 0, and thus,       $P(l^{\prime} \in \hat{\mathcal{A}}) \rightarrow 0$.

Now we want to prove the consistency of local selection only considering $l \in \mathcal{A}$ and $k \in \mathcal{B}$. Similarly, it suffices to show that for any $k^{\prime} \notin \mathcal{B}$, $P(k^{\prime} \in \hat{\mathcal{B}} | l \in \mathcal{A}) \rightarrow 0$.

If $k^{\prime} \in \hat{\mathcal{B}}$, under KKT conditions, we have
\begin{equation}
       2\boldsymbol{U}_{l}^{T}(\boldsymbol{r}_{(-l)} - \boldsymbol{U}_{l}\boldsymbol{b}_{l})I_{k^{\prime}} = n\Delta_{n}\lambda_{1}\hat{w}^{(1)}_{l}\text{sgn}(b_{lk^{\prime}}) + nM_{n}^{-1/2}\lambda_{2}^{2}\hat{w}^{(2)}_{l}b_{lk^{\prime}}/||\boldsymbol{b}_{l}||_{2},
       \label{eq:kkt_faddos2}
\end{equation}
where $I_{k^{\prime}}$ denotes $(M_{n}+d)$-dimensional unit vector with 1 at the $k^{\prime}$th entry. Following a similar reasoning as above, we show the probability that \eqref{eq:kkt_faddos2} holds tends to be 0 as well. Hence, $P(k^{\prime} \in \hat{\mathcal{B}} | l \in \hat{\mathcal{A}}) \rightarrow 0$.
\end{proof}

\newpage

\section{Simulation Studies}
\subsection{Data Generation}

\begin{figure}[b!]
\centering
\begin{subfigure}{1\textwidth}
  \centering
  \includegraphics[width=0.76\textwidth,height=0.38\textwidth]{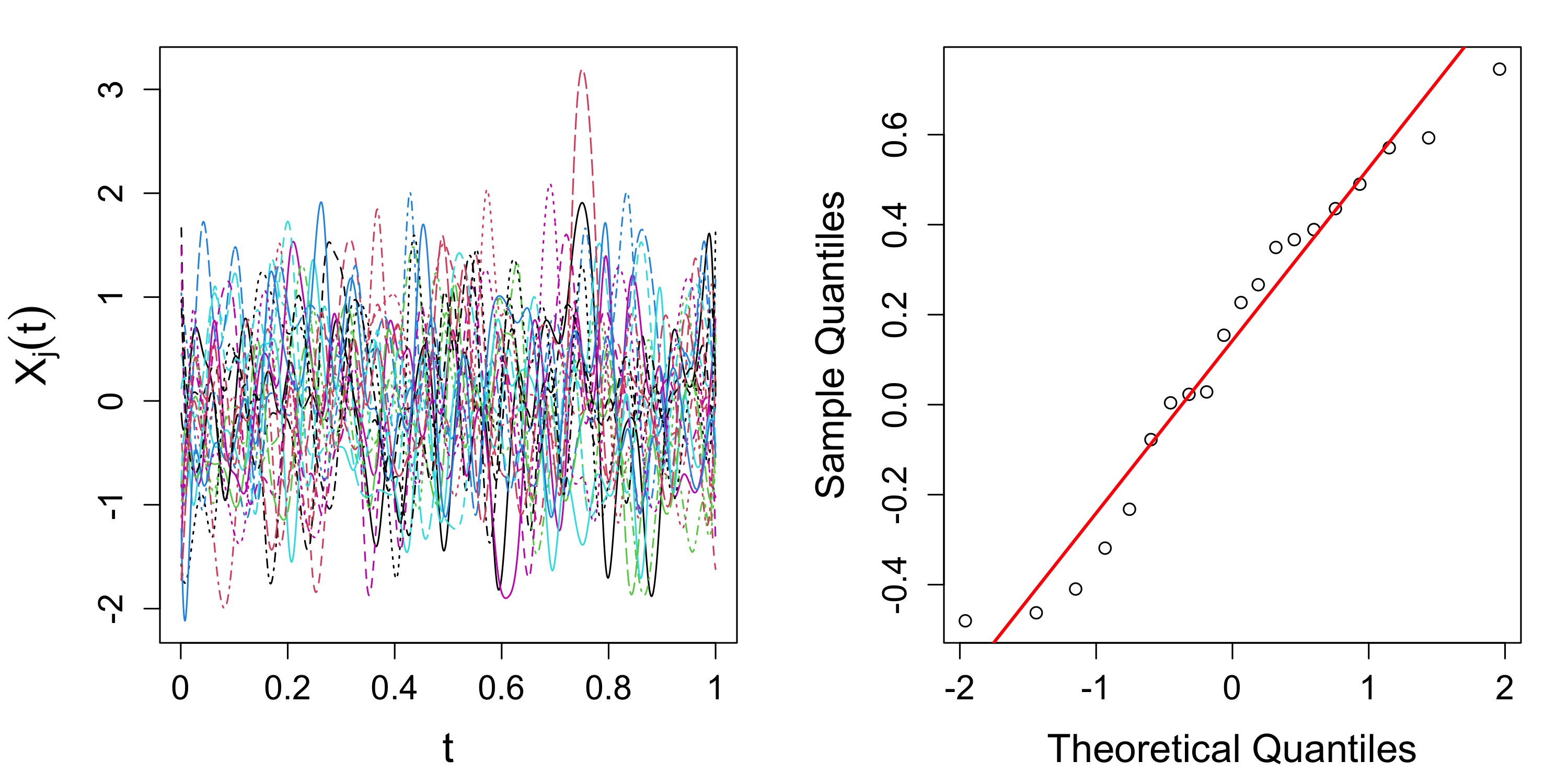}
  \caption{\hspace*{-5mm}}
\end{subfigure}
\begin{subfigure}{1\textwidth}
  \centering
  \includegraphics[width=0.9\textwidth,height=0.3\textwidth]{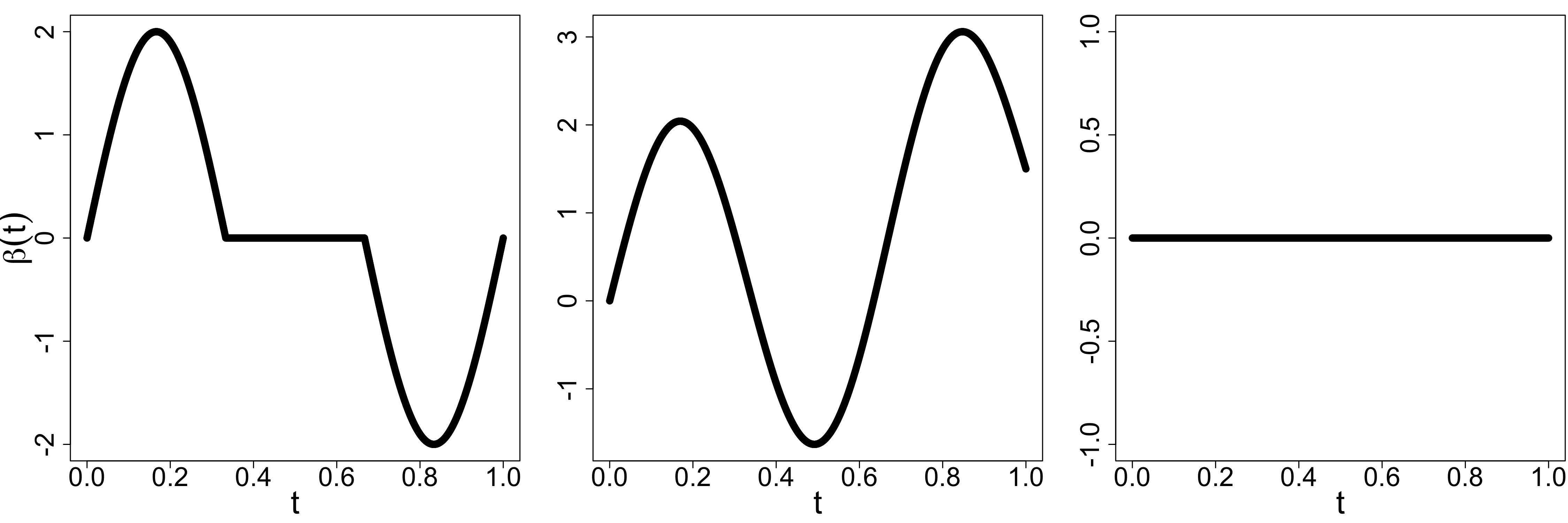}  
  \caption{\hspace*{-5mm}}
\end{subfigure}
\caption{Generating functional predictor, response, and coefficient functions in simulations study: (a) The left panel is generated $X_{ij}(t), j=1,\dots,10$ for all samples. Each curve represents the functional covariate for a subject over the time domain. The right panel is QQ-plot of generated $Y_{i}$. Both figures are drew under the sample size $n=20$; (b) True coefficient functions, from left to right panel, are $\beta_{1}(t)$, $\beta_{2}(t)$, and $\beta_{j}(t),j=3,\dots,10$.}
    \label{fig:true_covariate_beta}
\end{figure}

Figure \ref{fig:true_covariate_beta} illustrates the simulated data and three different types of coefficient functions. The functional predictors are generated to be standardized and independent, which ensures a fair assessment of method performance, without confounding factors of differing scales or inter-predictor dependencies. Additionally, the three types of coefficient functions demonstrate the global and local sparsity structure in the functional linear regression model.

\subsection{Effects of the Smoothness Parameters}

\begin{figure*}[htb!]
    \centering
     \includegraphics[width=0.9\textwidth, height=0.45\textwidth]{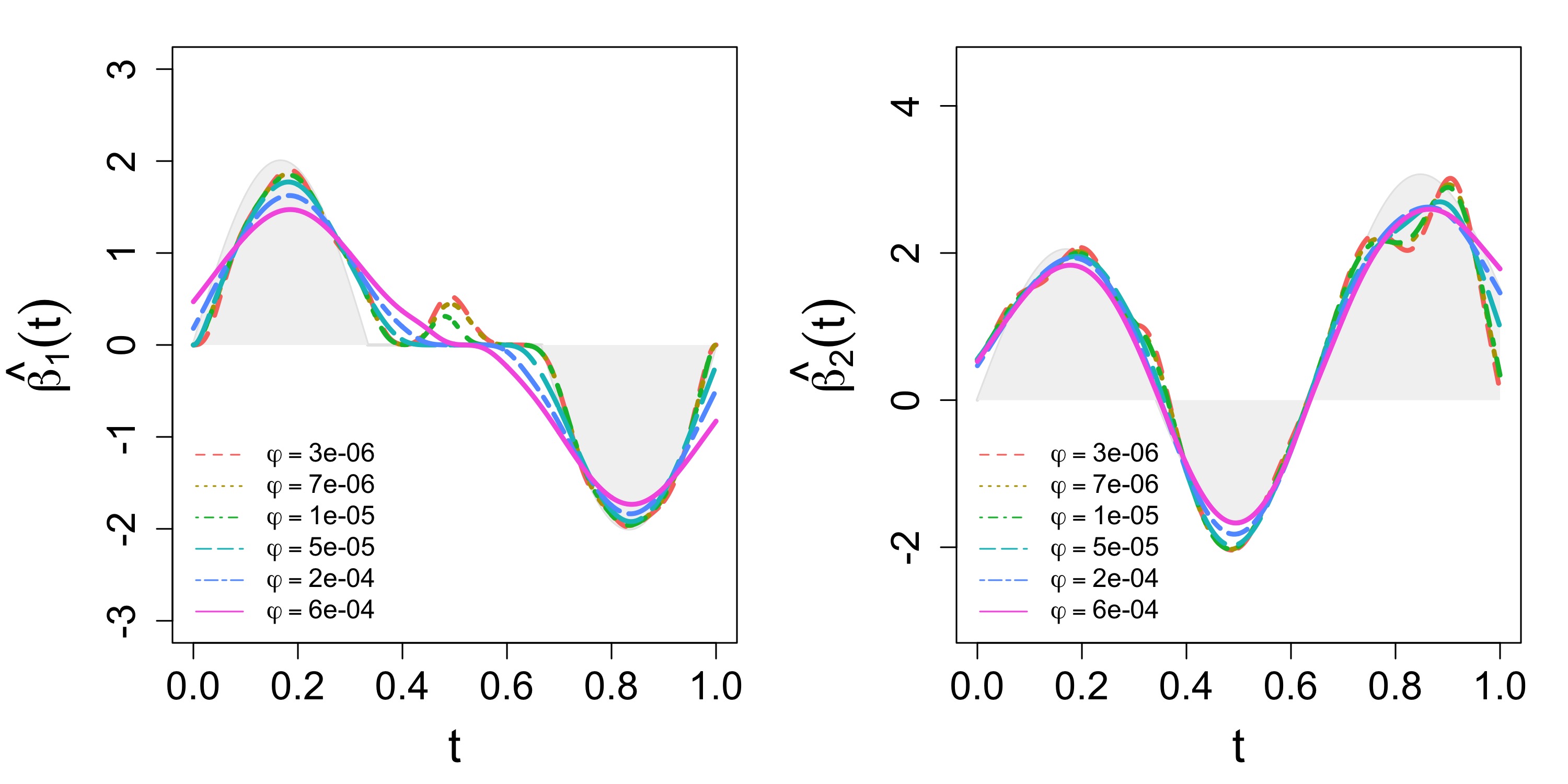}
    \caption{Estimated coefficient functions for $\beta_{1}(t)$ and $\beta_{2}(t)$ of the proposed estimator FadDoS with varying $\varphi$.}
    \label{fig:choose_phi}
\end{figure*}

While the functional $\ell_{1,2}$ penalty induces global sparsity, the smoothing parameter $\varphi$ controls overall curvature to ensure estimate smoothness. Figure \ref{fig:choose_phi} illustrates the effect of varying $\varphi$ with $\lambda_{1}$ and $\lambda_{2}$ fixed. Insufficiently small $\varphi$ leads to excessively wiggly functional estimates compared to the true functions. This impedes accurate identification of zero subregions and decreases the TNR due to inadequate shrinkage. Conversely, excessively large $\varphi$ overly linearizes the estimates. As shown in Table \ref{table:choose_phi}, moderate values of $\varphi$ around 5e-5 yield superior PMSE and ISE. The optimal level of smoothing avoids both under- and over-smoothing, thereby enabling precise estimation of zero subregions while maintaining estimate accuracy. Through controlling total curvature, the smoothing parameter provides localized regularization that complements the sparsity induced by the functional $\ell_{1,2}$ penalty.

\begin{table}[htb!]
    \centering  \hspace*{-3mm}
\scalebox{0.9}{
\begin{tabular}{|c c c c c c c|} 
\hline
  $\varphi$ & PMSE ($\times 10^{-2}$) & $\text{ISE}_{0}(\hat{\beta}_{1})$ & $\text{ISE}_{1}(\hat{\beta}_{1})$ &  $\text{ISE}(\hat{\beta}_{2})$ & $\sum_{j=3}^{10}\text{ISE}(\hat{\beta}_{j})$ & avgTNR \\[0.5ex]
\hline
 3e-6 & 2.59(0.18) & 52.87(53.87) & 103.83(48.22) & 99.52(36.11) & 0.37(3.29) & 0.99(0.0176)\\[0.5ex]
 7e-6 & 2.54(0.17) & 32.56(41.13) & 85.29(44.36) & 82.69(33.41) & 0.27(2.68) & 0.99(0.13)\\[0.5ex]
 1e-5 & 2.52(0.17) & 26.38(37.95) & 78.68(41.63) & 76.37(32.58) & 0.25(2.49) & 0.99(0.13) \\[0.5ex]
 5e-5 & 2.46(0.17) & 12.41(18.06) & 64.67(40.28) & 56.78(28.58) & 0.00(0.00) & 1.00(0.00) \\[0.5ex]
 2e-4 & 2.50(0.18) & 28.48(24.42) & 95.83(49.68) & 53.38(29.61) & 0.00(0.00) & 1.00(0.00)\\[0.5ex]
  6e-4 & 2.64(0.20) & 69.21(38.84) & 173.73(60.80) & 75.7(40.58) & 0.00(0.00) & 1.00(0.00)\\[0.5ex]
 \hline
        \end{tabular}}
        \caption{PMSE ($\times 10^{-2}$), ISE, and average TNR (avgTNR) of the proposed estimator FadDoS with varying $\varphi$. The test sample size is 1000. The entry in the parenthesis corresponds to the standard deviation among 100 simulation replicates.} 
        \label{table:choose_phi}
\end{table}



\clearpage

\section{Illustration of Timed Up and Go Test}

\begin{figure}[htb!]
\centering
\begin{subfigure}{.45\textwidth}
  \centering
  \includegraphics[width=0.84\textwidth,height=0.7\textwidth]{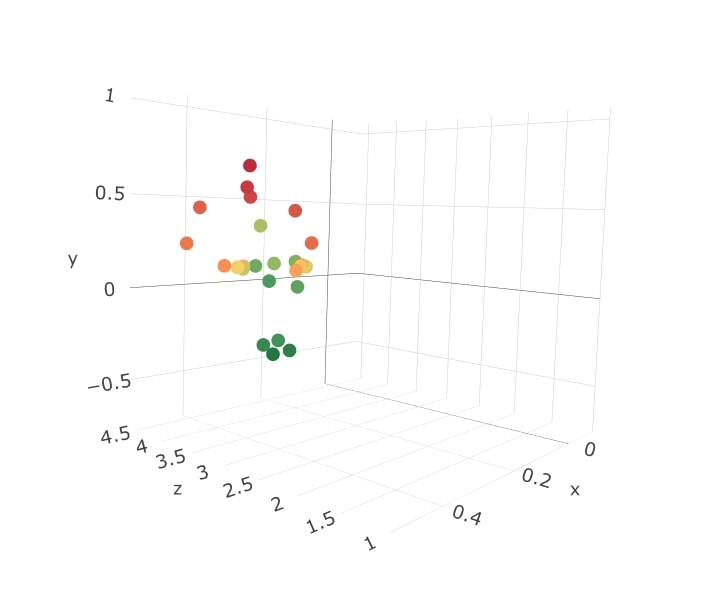}
  \caption{}
\end{subfigure}
\begin{subfigure}{.45\textwidth}
  \centering
  \includegraphics[width=0.84\textwidth,height=0.7\textwidth]{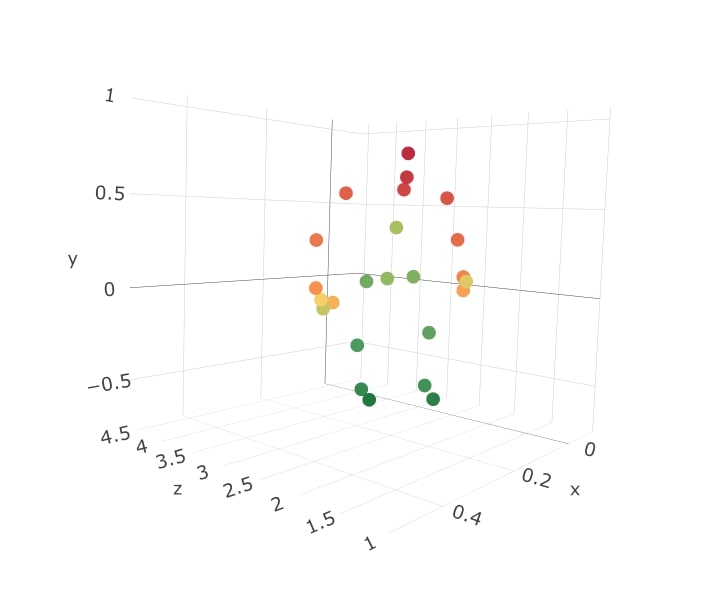}  
  \caption{}
\end{subfigure}
\begin{subfigure}{.45\textwidth}
  \centering
  \includegraphics[width=0.84\textwidth,height=0.7\textwidth]{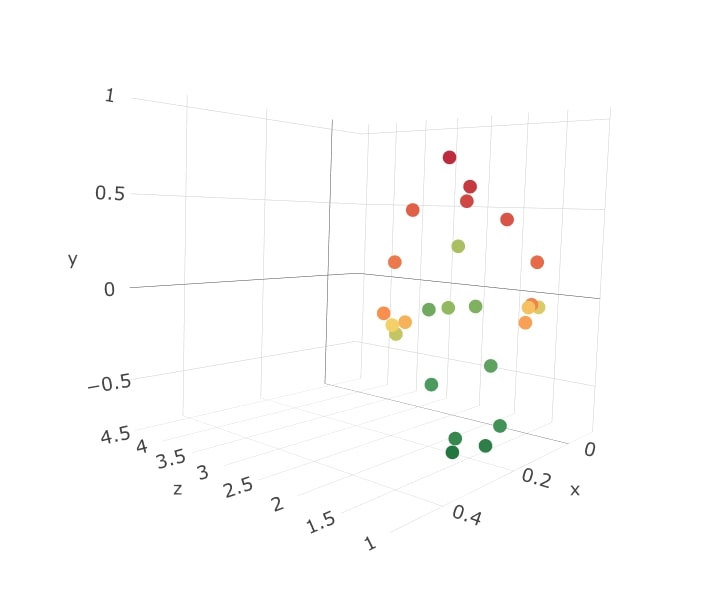}  
  \caption{}
\end{subfigure}
\begin{subfigure}{.45\textwidth}
  \centering
  \includegraphics[width=0.84\textwidth,height=0.7\textwidth]{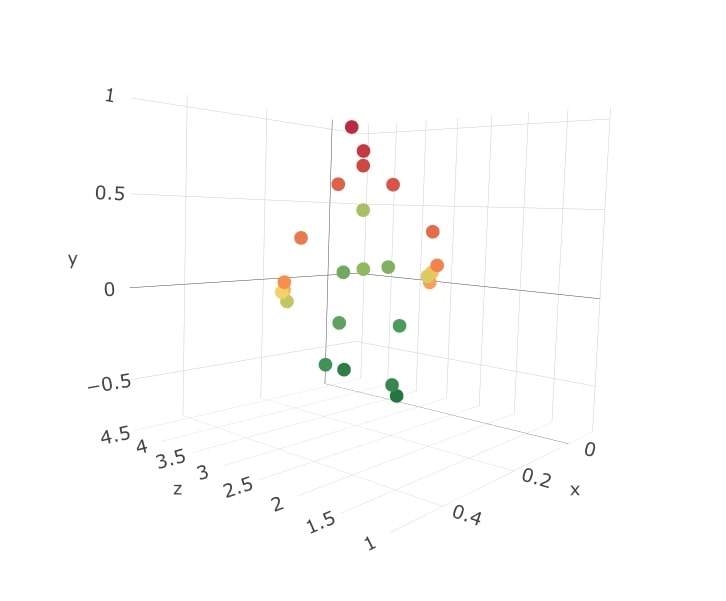}  
  \caption{}
\end{subfigure}
\begin{subfigure}{.45\textwidth}
  \includegraphics[width=0.84\textwidth,height=0.7\textwidth]{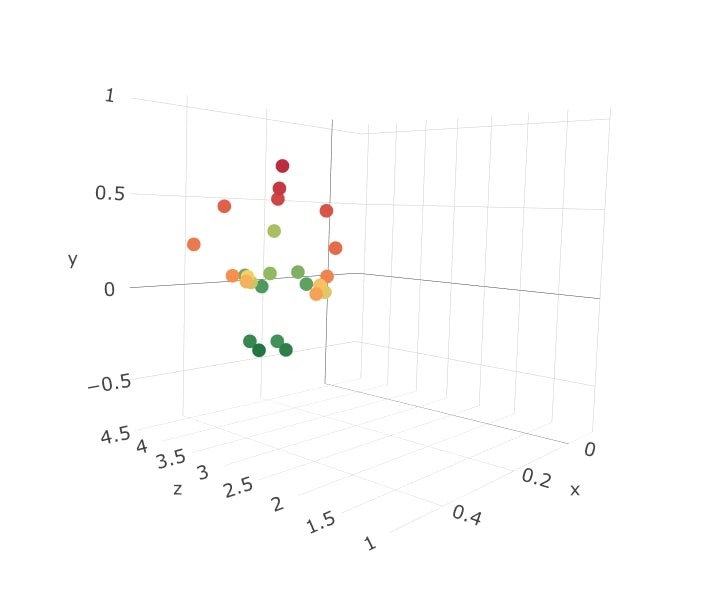}  
  \centering
  \caption{}
\end{subfigure}
\caption{Instruction of Timed Up and Go (TUG) Test : (a) Stand up from the chair; (b) Walk forward at a normal pace; (c) Turn; (d) Walk backward to the chair at a normal pace; (e) Sit down. Each joint is color coded as Figure 1(b).}
\label{fig:snapshot}
\end{figure}

\newpage

\bibliography{supp}